\documentclass[amsmath,amssymb,aps,twocolumn,superscriptaddress]{revtex4-2}
\usepackage{caption}
\usepackage[utf8]{inputenc}
\usepackage{graphicx}% Include figure files
\usepackage{dcolumn}% Align table columns on decimal point
\usepackage{bm}% bold math
\usepackage{siunitx}% try to write down numbers and units in a consistent way
\usepackage[version=4]{mhchem}% change $\mathrm{SO_2}$ or SiO${_2}$ to -> \ce{SiO2}
\usepackage{natbib}%customising references
\usepackage{float}%fix SI figures with specifier H
\usepackage[shortlabels]{enumitem}

\usepackage{xcolor}

\usepackage{mathtools}
\usepackage{braket}
\usepackage{amsmath}%
\usepackage{MnSymbol}%
\usepackage{wasysym}%
\usepackage{cancel}
\usepackage{physics}
\usepackage{caption}
\usepackage{subcaption}
\usepackage[english]{babel}
\usepackage[T1]{fontenc}
\usepackage{placeins}

\usepackage[colorlinks,citecolor=blue,urlcolor=red]{hyperref}
\usepackage{cleveref} 
\usepackage{chemformula}
\usepackage[newcommands]{ragged2e}

\newcommand{\red}[1]{{\color{red}#1}}

\makeatletter
\def\maketitle{
\@author@finish
\title@column\titleblock@produce
\suppressfloats[t]}
\makeatother
\newcommand{\beginsupplement}{
    \setcounter{table}{0}
    \renewcommand{\thetable}{S\arabic{table}}
    \setcounter{figure}{0}
    \renewcommand{\thefigure}{S\arabic{figure}}
    \setcounter{equation}{0}
    \setcounter{section}{0}
    \renewcommand{\theequation}{S\arabic{equation}}
}

\begin{document}

\title{Critical fluctuations in a confined driven-dissipative quantum condensate}

\author{Hassan Alnatah}
\thanks{Address correspondence to: haa108@pitt.edu}
\affiliation{Department of Physics, University of Pittsburgh, 3941 O’Hara Street, Pittsburgh, Pennsylvania 15218, USA}

\author{Paolo Comaron}
\affiliation{Department of Physics and Astronomy, University College London, Gower Street, London, WC1E 6BT, UK}

\author{Shouvik Mukherjee}
\affiliation{Joint Quantum Institute, University of Maryland and National Institute of Standards and Technology, College Park,
Maryland 20742, USA}

\author{Jonathan Beaumariage}
\affiliation{Department of Physics, University of Pittsburgh, 3941 O’Hara Street, Pittsburgh, Pennsylvania 15218, USA}

\author{Loren N. Pfeiffer}
\affiliation{Department of Electrical Engineering, Princeton University, Princeton, New Jersey 08544, USA}

\author{Ken West}
\affiliation{Department of Electrical Engineering, Princeton University, Princeton, New Jersey 08544, USA}

\author{Kirk Baldwin}
\affiliation{Department of Electrical Engineering, Princeton University, Princeton, New Jersey 08544, USA}

\author{Marzena Szymańska}
\affiliation{Department of Physics and Astronomy, University College London, Gower Street, London, WC1E 6BT, UK}

\author{David W. Snoke}
\affiliation{Department of Physics, University of Pittsburgh, 3941 O’Hara Street, Pittsburgh, Pennsylvania 15218, USA}

\date{\today}

\begin{abstract}

Phase fluctuations determine the low-energy properties of quantum condensates. 
However, at the condensation threshold, both density and phase fluctuations are relevant.
While strong emphasis has been given to the investigation of phase fluctuations, which dominate the physics of the quantum system away from the critical point---
number fluctuations have been much less explored, even in thermal equilibrium.
In this work, we report experimental observation and theoretical description of fluctuations in a circularly-confined non-equilibrium Bose-Einstein condensate of polaritons near the condensation threshold.
We observe critical fluctuations, which combine the number fluctuations of a single-mode condensate state and competition between different states. The latter are analogous to mode hopping in photon lasers. 
Our theoretical analysis indicates that this phenomenon is of a quantum character, while classical noise of the pump is not sufficient to explain the experiments.
The  manifestation of a critical quantum state competition unlocks new possibilities for the study of condensate formation while linking to practical realizations in photonic lasers.

\end{abstract}

\maketitle
\

\par
\section{INTRODUCTION}

\noindent
The experimental realization of Bose-Einstein Condensates (BECs) enabled investigations of macroscopic quantum systems. One fundamental question relates to the role of fluctuations close to the critical point of a BEC transition. 
Critical quantum fluctuations arise from the uncertainty principle and play an essential role in a wide range of physical phenomena such as the structure of the universe, the Casimir effect and transitions between two competing phases of matter \cite{white1994anisotropies,casimir1948attraction,gambassi2009casimir,chakravarty1989two,ruegg2008quantum,schroder2000onset,chomaz2016quantum,papp2008bragg}. Understanding such fluctuations is important for  statistical physics of phase transitions and quantum critical phenomena.

Up to now, the focus has largely been on phase fluctuations, which are fundamental for understanding the physics away from the critical point \cite{das2019critical}.
Yet, at the threshold of condensation, density fluctuations play a crucial role, while they have been much less explored.
In the context of {\textit{equilibrium}}  systems, number fluctuations have been characterised to some extent in dilute atomic gases \cite{Giorgini1998,Meier1998,Kristensen2019} and photon in a dye \cite{Schmitt2014,Ozturk2019} condensates. 
For the {\textit{out-of-equilibrium}} systems, however, the physics of  critical number fluctuations has not been explored yet.
Here the scenario is more complicated; in addition to the usual density fluctuations in a single lowest energy state, we can witness a competition between different modes in the condensate formation process, which is particularly pronounced at  vicinity of the critical point, where the condensate occupation is low. 

\par
Due to their strong nonlinear properties, microcavity exciton-polaritons (called here simply ``polaritons'') are the perfect candidates for the investigation of critical density fluctuations in non-equilibrium quantum systems.
They are quasiparticles arising from the strong coupling of cavity photons and excitonic transitions, typically in semiconductor quantum wells (QWs) placed at the antinodes of an optical cavity. Polaritons can be viewed as photons with nonlinear interactions, or alternatively as excitons with very light effective mass (typically $10^{-4}$ times the mass of an electron in vacuum). As interacting bosons, polaritons demonstrate quantum phenomena such as Bose-Einstein condensation ~\cite{kasprzak2006bose,balili2007bose}, superfluidity~\cite{amo2009superfluidity} and quantized vortices~\cite{lagoudakis2008quantized,lagoudakis2009observation,sanvitto2010persistent,nardin2011hydrodynamic,tosi2012geometrically,liu2015new} at temperatures from tens of Kelvin up to room temperatures.

\par
In the present work, we undertake a detailed experimental and theoretical investigation of fluctuations near the condensation transition in a non-equilibrium system of
non-resonantly pumped polaritons placed in an annular trap geometry.
We explore theoretically the effect of quantum and classical noise on the system behaviour, and in particular on the mechanisms of the density fluctuations and mode competition.
Apart from the fundamental interest, our results are also relevant for practical design of photonic lasers, where multimode switching ---originating from their weakly-interacting nature--- leads to  ``telegraphic noise"~\cite{degen1999transverse,zhang2012towards,deng1997radiation,pereira1998pinning} instabilities.
% %

\begin{figure*}
\centering
\includegraphics[width=0.85\textwidth]{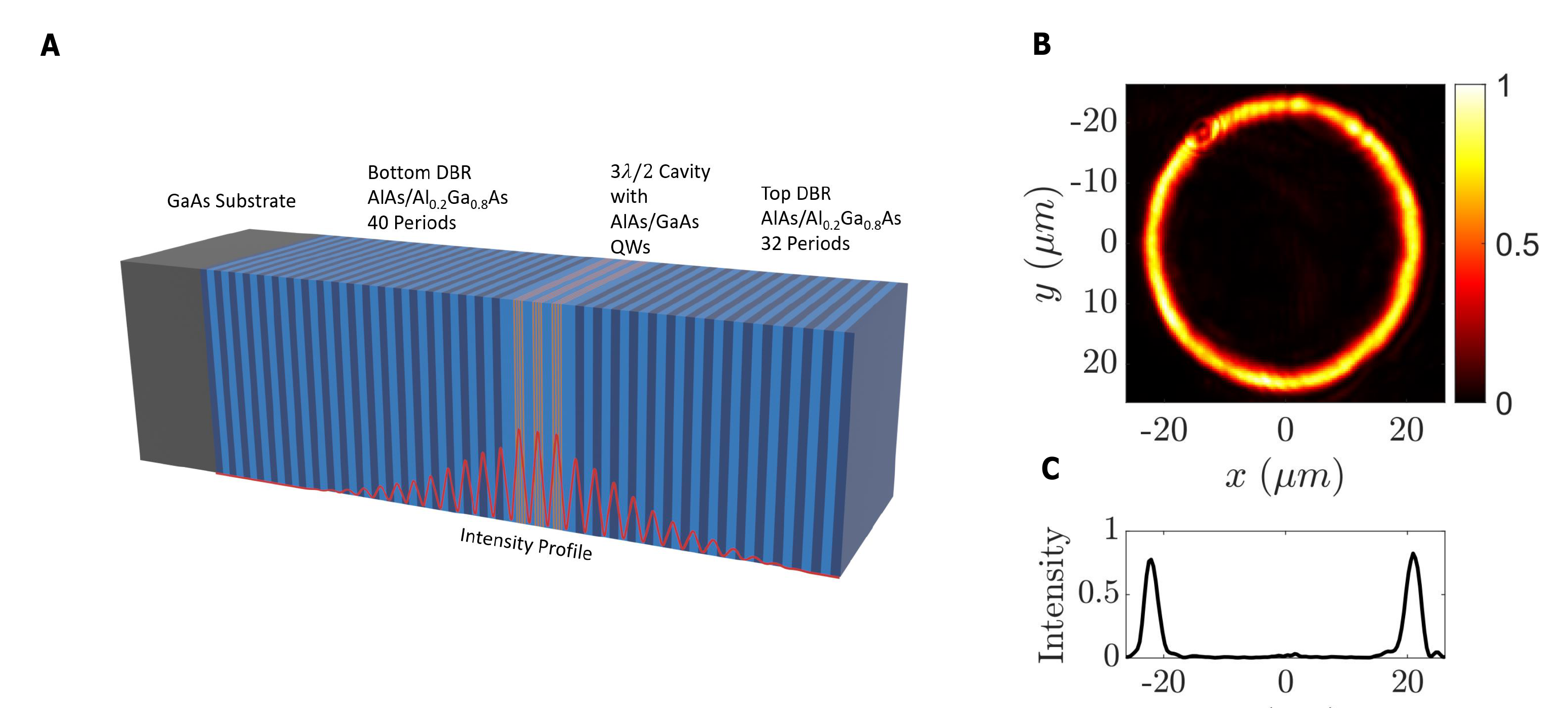}
\caption{
\textbf{Optical trapping using annular pump in a GaAs microcavity.} 
\textbf{(A)} Microcavity structure used in the experiment. The dark and light blue layers represent the DBRs used to confine the light field. The yellow layer indicates the quantum wells, where bound excitons are formed and the red lines indicate the intensity of the
confined optical field. \textbf{(B)} Normalized real-space image of the excitation laser generated by a spatial light modulator. \textbf{(C)} A slice of the intensity of the excitation ring at y=0 line through the center of the excitation ring pattern shown in \textbf{(B)}.
}
\label{fig1}
\end{figure*}

%%%%%%%%%%%%%%%%%%%%%%%%%%%%%%%%%%%%%%%%%%%%%%%%%%%%%%
%%%%%%%%%%%%%%%%%%%%%%%%%%%%%%%%%%%%%%%%%%%%%%%%%%%%%%
%

\

\section{RESULTS AND DISCUSSION}

\begin{figure*}
\includegraphics[width=\textwidth]{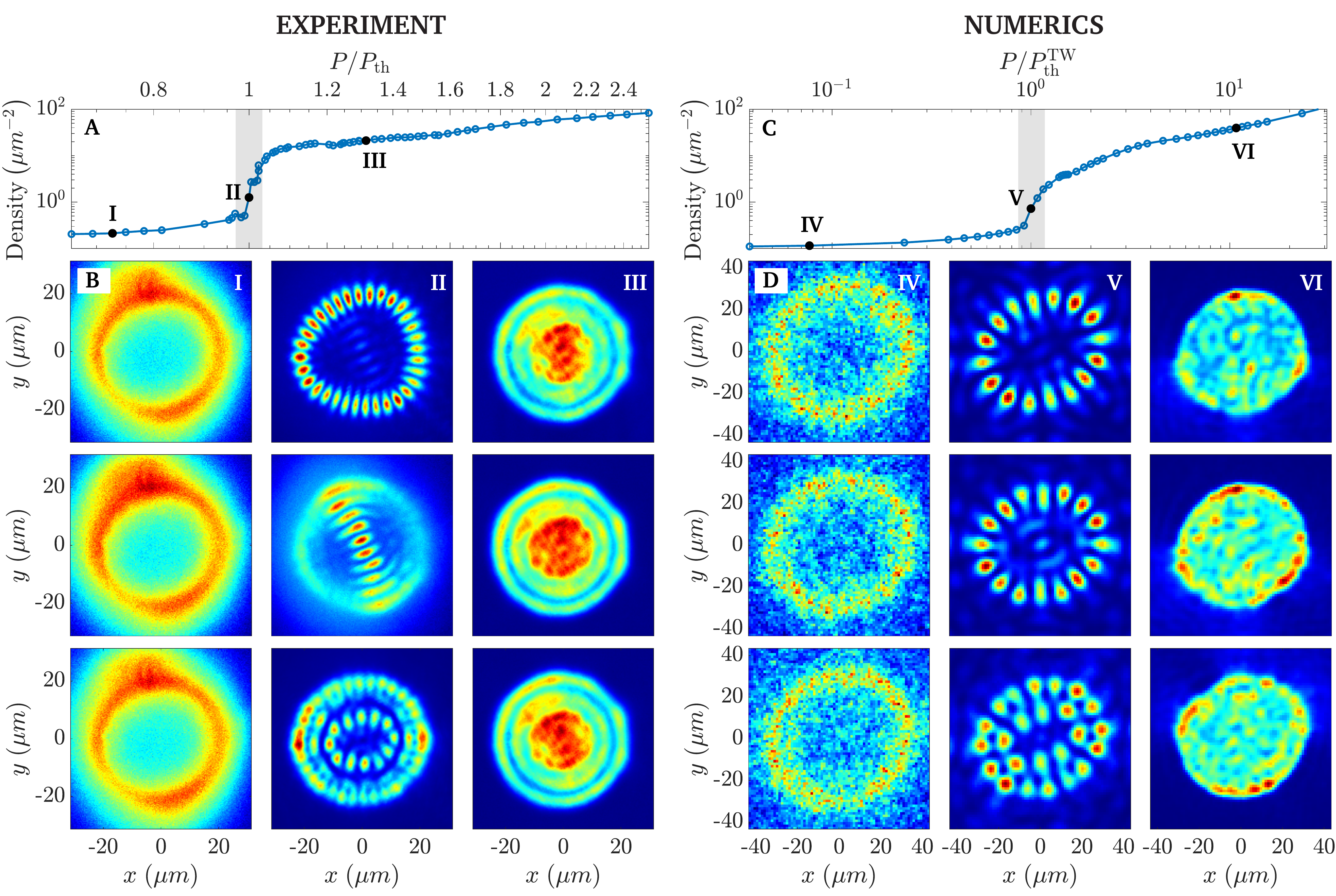}
\caption{\textbf{Critical mode competition  and density patterns}
in experiment (Left panels \textbf{(A,B)} and theoretical modeling (Right panels \textbf{(C,D)}). Panels \textbf{(B,D)} are selected snapshots of the polariton density distributions, where each column corresponds to a fixed pump power below (I column), near (II column) and well above (III column) the condensation threshold. The exact $P/P_{\text{th}}$ for each column is indicated by the black circles in \textbf{(A)} and \textbf{(C)}.
In panel \textbf{(D)}, the single-realization density profiles are integrated in time after the system has reached the non-equilibrium steady state with the integration period of $\tau_\mathrm{int} = 20 \mathrm{ns}$.
}
\label{fig2}
\end{figure*}

\par

\noindent
In order to investigate the fluctuations close to the critical point for condensation, we confined the polariton fluid in a circular trap, using a spatial light modulator to focus the pump laser in a ring, as shown in Figure (Fig.~\ref{fig1}(\textbf{B})). The pump laser acted as an incoherent source of excitons and polaritons, since it had photon energy well above the exciton energy, so that many phonons had to be emitted before the hot carriers cooled and turned into excitons and polaritons. This incoherent pumping therefore had two effects: it created both the potential energy profile of the trap, due to the repulsion of polaritons from the slow-moving excitons in the pump region, and it also acted as the source of the polaritons, as excitons converted into polaritons. 
As the pump power increased, the system went from non-condensed to a single-mode condensed regime, as reported in previous works.
The large number of DBR periods gives the cavity a high Q-factor, resulting in a cavity lifetime of $\sim$135 ps and a polariton life time of $\sim$270 ps at resonance~\cite{steger2013long,steger2015slow,steger2012single}
(see Materials and Methods for a detailed description).
\par
The sample was cooled in a continuous-flow cold-finger cryostat, which was held at a temperature of $\sim$5 K. To produce polaritons, the sample was excited non-resonantly with a continuous-wave (cw) laser, which was modulated by an optical chopper at 404 Hz with a duty cycle of 1.7\% to prevent sample heating. The ring-shaped trap was generated by modulating the phase front of the laser using a spatial light modulator (SLM). 
\par
The diameter of the ring-shaped optical trap used in this manuscript was 45 $\mu$m, which is shown in Figure.~\ref{fig1}\textbf{(B)}. The cavity detuning from the exciton resonance was $\delta = -10.1 \;\text{meV}$, corresponding to an exciton fraction $\left | X \right |^{2} = 0.3$ for the lower polariton. The photoluminescence (PL) was collected using a microscope objective with a numerical aperture (NA) of 0.75, and was imaged onto the entrance slit of a spectrometer. The image was then sent through
the spectrometer to a charged coupled device (CCD) for time integrated imaging. 
\par

Polariton condensates allow the study of a range of behavior as the detuning, and consequently the photon fraction of the polaritons, is changed. In the limit of high exciton fraction, the modes are closely spaced together in a near-continuum, and the interactions between the particles are strong, so that the dynamics are those of a close to equilibrium Bose-Einstein condensate \cite{sun2017bose,caputo2018topological}. As the exciton fraction is decreased, the interactions become weaker and their mass lighter, so that they approach behavior that is more like a standard laser. The experiments we have performed here are at a middle ground; the polaritons are in the strong coupling limit, and travel tens of microns away from the pump region, but have resolvable transverse modes similar to those of a VCSEL.

\par
We recorded several snapshots of the photoluminescence (PL) emission patterns from the trap for different pump powers. Each snapshot was time integrated for 100 ms. Since the pump laser was optically chopped, each snapshot is a sum of 42 images from separate laser pulses, each of which was about $42 \mu s$ long. Crucially, the pump conditions are identical for all snapshots at each pump power, since we used a stabilized M Squared laser, with very low noise.
Fig.~\ref{fig2}\textbf{(A)} shows the density across the non-condensed-to-condensed transition as the pump power was increased; in
Fig.~\ref{fig2}\textbf{(B)} we show different snapshots of the PL in real space below, at and above  threshold for condensation. Below threshold, polaritons remain in the vicinity of the pump spot and the PL looks spatially similar from one snapshot to the next as shown in Fig.~\ref{fig2}\textbf{(B)}, column I. For pump power above the condensation threshold, the polariton population builds up inside the trap. When the density becomes larger than the critical density,  stimulated scattering becomes significantly enhanced, giving rise to a condensate mode. Such modes have been seen in previous experiments with photonic detuning \cite{sun2018stable,dreismann2014coupled,askitopoulos2015robust,manni2011spontaneous,cristofolini2013optical,ouellet2016spatial}, and resemble pure lasing modes in vertical cavity, surface-emitting lasers (VCSELs) with imposed circular symmetry \cite{degen1999transverse,zhang2012towards,deng1997radiation,pereira1998pinning}. 
\par
As expected, near the onset of condensation, we observe strong density fluctuations. Some of these fluctuations can be attributed to hopping of the condensate between different discrete eigenstates. Even though the pump conditions are identical from one snapshot to the next, the polariton gas near the threshold for condensation does not always condense to the same mode, but rather switches  between several distinguishable modes as illustrated in Fig.~\ref{fig2}\textbf{(B)}, column II. 
Interestingly, although each snapshot is the result of summing 42 images from separate laser pulses, statistical clumping of certain modes ---in small samples of 42--- is able to reproduce the same mode switching  effect (see Supplementary Information for more details). Notably, unlike optical multistability or bistability mechanisms, which can take place under resonant  excitations at higher pump powers~\cite{PhysRevA.69.023809,10.1063/5.0136380}, the critical fluctuations we see are only present at the condensation threshold.
Surprisingly, we find that the time scale of the fluctuations is very long compared to the intrinsic scattering time scale of the polaritons; features are seen to differ in images recorded with time integrations of 100 ms.
This indicates that the system has metastable telegraph-type switching between modes, as have been seen for lasers \cite{ohtsu1985precise,ohtsu1986analyses,ohtsu1989analyses}. Here, this corresponds to the mode competition in the condensate formation process. In equilibrium BEC, it is the ground state of the system, which becomes macroscopically occupied. Since our system is driven-dissipative, and out-of-equilibrium, the process of the condensate formation is more complex. Which mode wins out to condense depends on the subtle interplay of gain and interactions effects. 
Eventually, as the pump power is increased away from the condensation threshold, the density fluctuations become less relevant and the polariton fluid occupies one mode for all different snapshots (see Fig.~\ref{fig2}\textbf{(B)}, column III).

\begin{figure*}
\includegraphics[width=\textwidth]{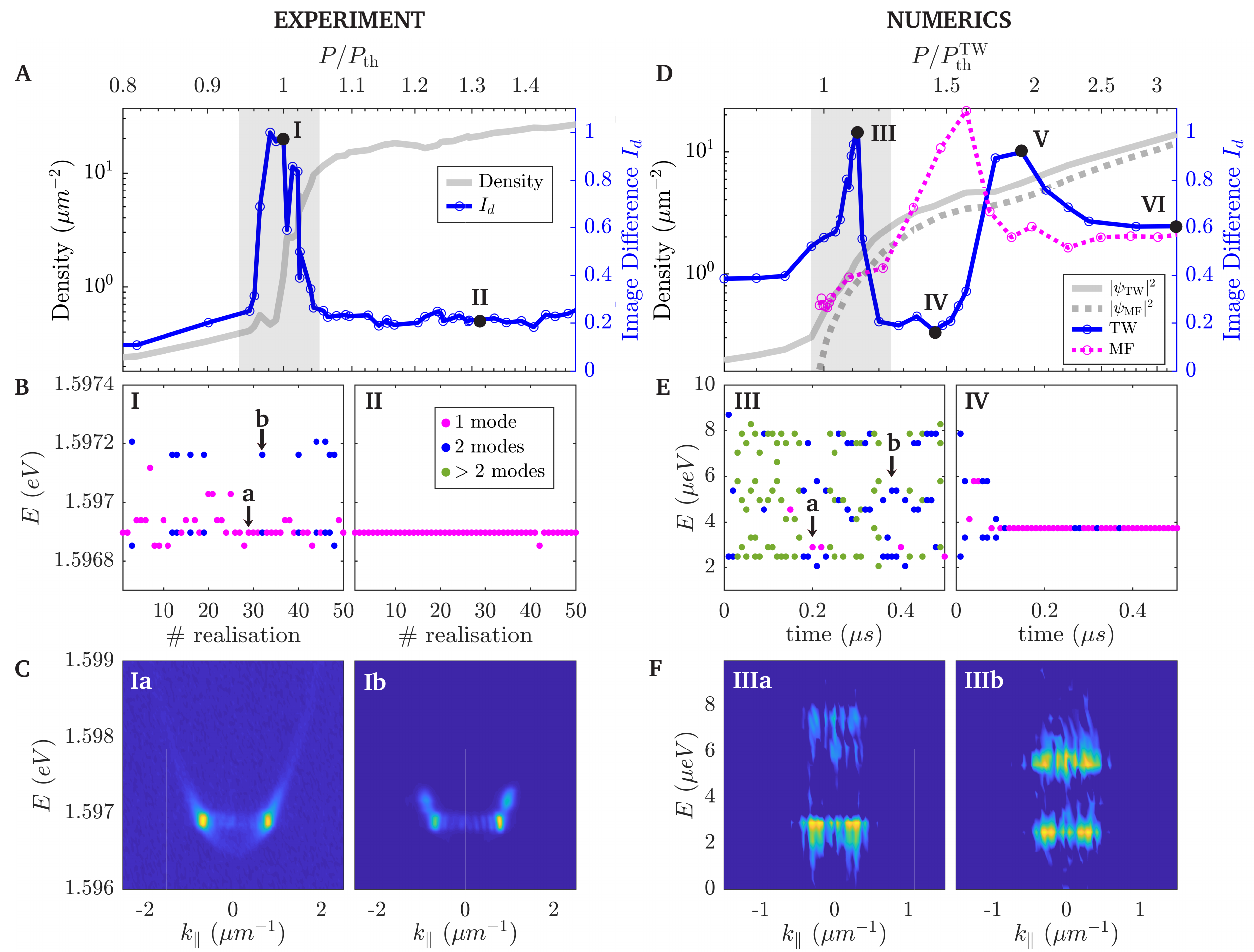}
\centering
\caption{\label{fig3}
\textbf{Image difference, energy hopping and spectra} from
experiment  (left\textbf{(A,B,C)})  and theoretical modelling (right \textbf{(D,E,F)} panels).
\textbf{(A)} Normalized average image difference $I_d$ given by Eq.~\ref{eq:image_diff} (blue circle and blue solid curve) and the polariton density (gray solid line) as a function of $P/P_{th}$. 
\textbf{(B)} The extracted energy peaks for each snapshot, ({I}) near and ({II}) far from the critical point. 
In some cases the condensate is single-mode (purple dots), while in others is multi-mode (blue dots). 
\textbf{(C)} Representative $I\left( k,E \right)$ snapshots from which the energy peaks were extracted. We show two examples for a single ({Ia}) and a multi-mode state ({Ib}) close to criticality. 
\textbf{(D)} The same as in panel \textbf{(A)}, but from theoretical mean-field (MF) and Truncated-Wigner (TW) analysis. 
Polariton densities are plotted with dashed (solid) gray line for MF (TW). The average image difference $I_d$ is shown as magenta (blue) circles and magenta (blue) solid line for MF (TW). 
In panels \textbf{(A)} and \textbf{(D)} the critical region is marked as a gray shaded area; consistent with previous works~\cite{Comaron2021}.  The threshold power was identified where the plot changes slope in a log-log scale.
\textbf{(E)} The extracted energy peaks along the time evolution of a single realization of the model (\ref{eq:GPE_pol}-\ref{eq:GPE_res}), within ({III}) and above ({IV}) the critical region. 
\textbf{(F)} The same as panel \textbf{(C)}, but for the theoretical analysis showing two time-integrated snapshots with different number of modes.
}
\end{figure*}

\

%%%%%%%%%%%%%%%%%%%%%%%%%%%%%%%%%%%%%%%%%%%%%%%%%%%%%%
%%%%%%%%%%%%%%%%%%%%%%%%%%%%%%%%%%%%%%%%%%%%%%%%%%%%%%
\subsection{Theoretical Modelling}
\noindent
Near the critical threshold for condensation, mean-field theories are known to become inaccurate since the fluctuations of the order parameter about its mean value become large, and the state of the many-body system is no longer accurately described by a single-particle wave function. 
In order to model the experimental results, 
we describe the strongly fluctuating polariton field  within the truncated Wigner (TW) approximation~\cite{carusotto2013quantum,estrecho2018single} in terms of a stochastic equation coupled to a rate equation for the excitonic reservoir.
The full model and further details regarding numerical methods can be found in the section Materials and Methods.

\par
 We solve the stochastic equations with a Wiener noise at each time step for many realizations of the initial noise. Both initial noise and the noise during the time evolution simulate quantum fluctuations. We emphasize that the amplitude of the Wiener noise in the system is not tuned externally, but rather, it is introduced self-consistently in the model. 
In order to simulate the experimental measurements, the polariton dynamics is left to evolve for a long time (up to $0.5~ \mu s$) after the Non-Equilibrium Steady-State (NESS) is reached. Then, each realization is independently integrated over different time-windows of the NESS.
In Fig. \ref{fig2} we report typical  results from our model.
While the behaviour of the polariton density is depicted in Fig.~\ref{fig2}\textbf{(C)}, typical steady-state realizations are plotted in Fig.~\ref{fig2}\textbf{(D)}, for three typical cases: below, near and above the critical point (Fig.~\ref{fig2}\textbf{(D)}, column IV, V and VI respectively).
Similar to the experimental cases explored in Fig.~\ref{fig2}\textbf{(A-B)}, below and far above threshold of condensation, the quantum noise plays no role, producing an ensemble of similar images for each realization (Fig.~\ref{fig2}\textbf{(C)}, columns IV and VI). 
Close to threshold, however, the quantum noise becomes important, giving rise to realizations with different modes (Fig.~\ref{fig2}\textbf{(D)}, column V). 

\

%%%%%%%%%%%%%%%%%%%%%%%%%%%%%%%%%%%%%%%%%%%%%%%%%%%%%%
%%%%%%%%%%%%%%%%%%%%%%%%%%%%%%%%%%%%%%%%%%%%%%%%%%%%%%
\subsection{Measuring critical mode competition}
\noindent
To gain a quantitative description of the critical fluctuation features exhibited by the polariton system, we calculated the normalized average image difference between different consecutive images from the experiments for a given set of parameters such as pump power. This can be written as
\begin{equation}
I_d=\frac{\left\langle R M S\left(I_{n+1}-I_n\right)\right\rangle}{\left\langle I_n\right\rangle}
\label{eq:image_diff}
\end{equation}
where $I_n$ is the $n$th intensity image in real space and RMS stands for the root mean squared. This average image difference is then normalized for each pump power by the average intensity for that set.
We plot this quantity for both the experiment and the numerical simulations in Fig.~\ref{fig3}\textbf{(A)} and \textbf{(D)}, respectively. 
While relatively small below and above the critical point, the average image difference $I_d$ is found to show a maximum exactly at the threshold region indicating that fluctuations are large near the condensation threshold. 
In order to verify that this behavior is due to the onset of competition between different modes, we make use of angle-resolved imaging to access the information in momentum space~\footnote{Similar to the real space images, the angle-resolved images are a sum of 42 laser pulses, each time integrated for 100 ms.}.  
Such a procedure allows us to extract the number of modes, together with their energy, over the whole set of images. 
In Fig.~\ref{fig3}\textbf{(B)} we plot energy hopping for powers at, and above, the critical threshold, revealing the single-mode to many-mode transition. 
The quantity $I(k_\parallel,E)$ 
is shown in Fig.~\ref{fig3}\textbf{(C)} for two typical configurations: single-mode (panel Ia) and multi-mode condensate (panel Ib). 

Our numerical simulations are in good agreement with the experiment; the numerical average image difference has a maximum near the threshold (point III in Fig.~\ref{fig3}\textbf{(D)}), while  it drastically decreases at lower and higher pump powers (point IV).
We can extract the different mode energies by Fourier transforming the computed wavefunction $\psi = \psi(\textbf{r},t)$ to $\psi = \psi(\textbf{k},\omega)$. 
Experimentally, we did not measure the energy fluctuations in real time. Instead, we extracted the mode energy for each snapshot, allowing us to plot the energy as a function of snapshot number. 
We can, however, access this information from our numerical simulations; transforming over different temporal windows, we extract the energy peaks for each spectra and plot the energy as a function of time (Fig.~\ref{fig3}\textbf{(E)}). 

\

%%%%%%%%%%%%%%%%%%%%%%%%%%%%%%%%%%%%%%%%%
%%%%%%%%%%%%%%%%%%%%%%%%%%%%%%%%%%%%%%%%%
\subsection{Mode hopping during relaxation}

\begin{figure}
\includegraphics[width=0.5\textwidth]{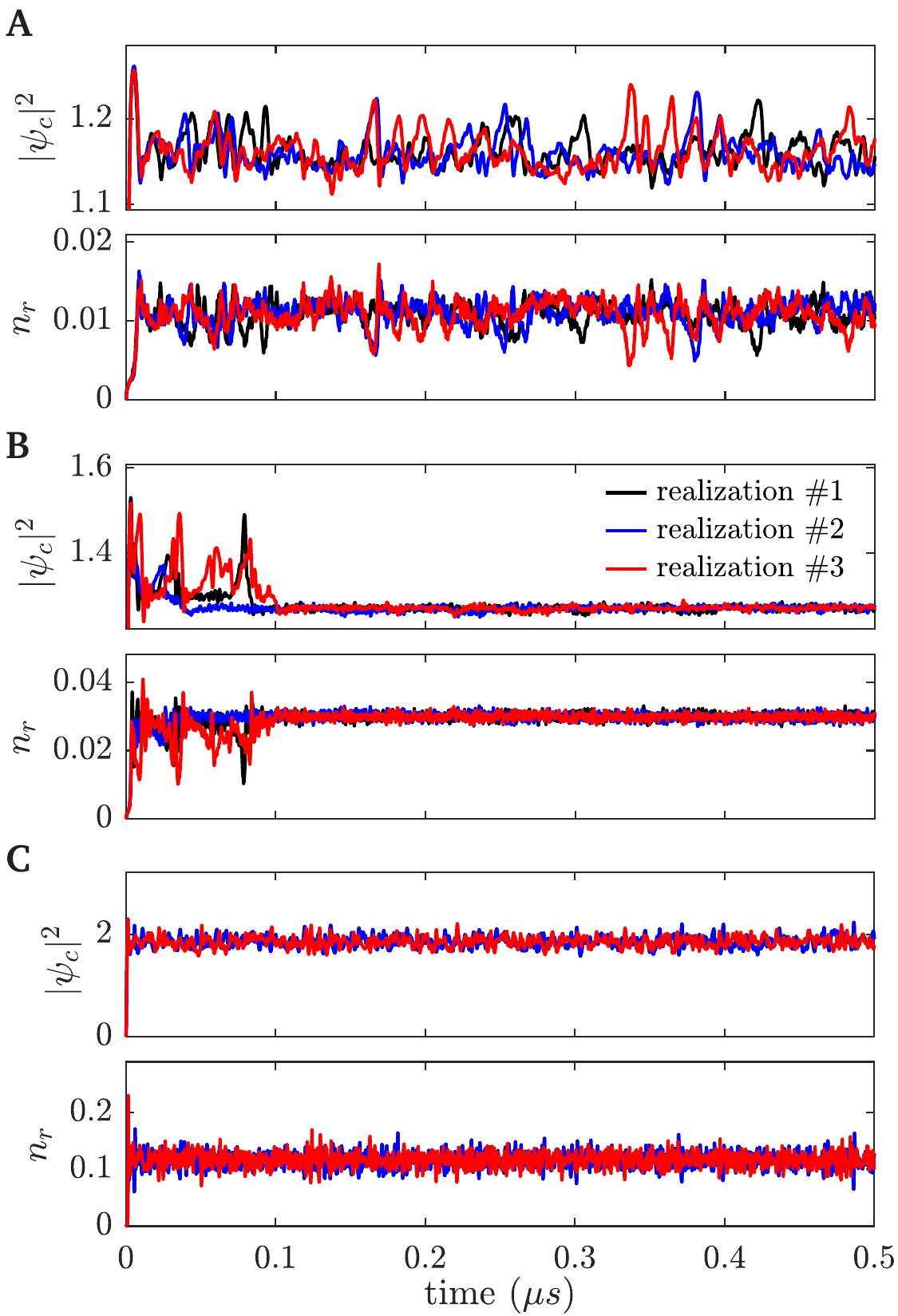}
\centering
\caption{
\textbf{Density evolution.} The simulated total condensate  $|\psi_c(t)|^2$ and reservoir $n_r(t)^2$ densities for \textbf{(A)} $P/P_\mathrm{th}^\mathrm{TW}{} = 1.12$, \textbf{(B)} $P/P_\mathrm{th}^\mathrm{TW}{} = 1.44$ and \textbf{(C)}, $P/P_\mathrm{th}^\mathrm{TW}{} = 3.2$, corresponding to  {III}, {IV} and {VI} of Fig.~\ref{fig3}\textbf{(D)}. The different curves  (black, blue and red) represent three different realizations. Density is in $\mu m^{-2}$ units. 
\label{fig4}}
\end{figure}

\noindent
Let us now focus on the mode hopping behaviour of the different cases shown in Fig.~\ref{fig3}\textbf{(E)}.
At pump powers well above threshold (Fig.~\ref{fig3}\textbf{(E)}, panel IV), the system undergoes energy fluctuations and hopping during only the early dynamics of the system, and then settles down to a single mode. 
Such a behaviour is linked to the relaxation dynamics expected to take place in coupled condensate-reservoir systems. 
In previous works, the relaxation dynamics and the dynamical instabilities have been found to be a direct effect of the polariton-reservoir repulsive interaction~\cite{Bobrovska2015,opala2018,degiorgi2014,Baboux,nalitov2019}, the latter being closely related to the polariton hole-burning effect~\cite{estrecho2018single}. 
In Fig.~\ref{fig4}\textbf{(B)}, corresponding to the case above threshold (point IV in Fig.~\ref{fig3}\textbf{(D)}), the strong density fluctuations take place up to times  $20 < \tau_\mathrm{RD}<100$ ns before eventually relaxing down.
It is interesting to note that at higher pump powers (Fig.~\ref{fig4}\textbf{(C)}), corresponding to point VI in Fig.~\ref{fig3}\textbf{(D)},
the NESS is reached after a very short relaxation dynamics with characteristic time of the order of a few nanoseconds. 

\

%%%%%%%%%%%%%%%%%%%%%%%%%%%%%%%%%%%%%%
%%%%%%%%%%%%%%%%%%%%%%%%%%%%%%%%%%%%%%
\subsection{Critical  mode competition}

\noindent
At pump powers close to the condensation threshold the situation  changes substantially.  
As depicted in Fig.~\ref{fig3}\textbf{(E)}, panel III the system   spends time in three or more distinguishable states along its entire dynamics.
Fig.~\ref{fig4}\textbf{(A)} shows that the mode competition leads to to strong density fluctuations.
In Fig.~\ref{fig3}\textbf{(F)} we plot the $I(k,E)$ spectra of two typical  modes (further details on the spectral analysis of the mode competition are reported in the Supplementary Information).
This numerical observation helps us to give a quantitative picture of the system's behavior at the critical point: different stochastic realizations (initialized with different numerical seeds) lead to a very different and independent dynamics.
Each stochastic path exhibits strong fluctuations in density and  hopping between different modes and energies.  
It is important to note here that such fluctuations persist for timescales up to $\sim 20$ns, which are orders of magnitude larger than the polariton lifetime, leading to persistent “intermittent” oscillations, which resemble telegraphic behavior observed in photonic systems~\cite{degen1999transverse,zhang2012towards,deng1997radiation,pereira1998pinning}.
Importantly, we note that, as opposed to the configurations at larger pumps discussed in the previous section, the fluctuations persist over the whole NESS dynamics.

While one would expect that by time-averaging over long periods, such fluctuations would eventually be washed out, we find that imaging of mode competition  is accessible even when integrating up to times comparable to the NESS duration (see Fig.~S6 and discussion in the SI).
The mode-competition effect can therefore be measured by making use of such ``statistical clumping'' of certain modes in small ensembles of measures; we discuss this in more detail in the Supplemental Information.
We also investigated the effects of disorder on the mode hopping. Introducing a static disorder profile, with amplitude and spatial correlations matching the experiment, we observe an enhancement of the image difference $I_d$. A more detailed discussion is reported in the Supplemental Information.

Finally, we note that in the experiment the measured time scales, of the order of a few microseconds, are longer than the ones observed in the theoretical analysis. 
We find that such a discrepancy can be explained by introducing low frequency  pump 
modulations, mimicking external perturbations that are intrinsic to the experimental setup, but not accounted in the earlier theoretical modeling. 
We analysed how such time scales depend on external perturbations; our results are discussed in the Supplementary Information.
We introduced small periodic modulations to the pump, of the order of the timescale of the numerical mode hopping ($\sim 20 \mathrm{ns}$).
We observed that the switching time scales can be extended up to the total duration of the numerical dynamics, i.e. on the order of microseconds, therefore matching the switching timescales of the experimental measurement. 

\

\subsection{Quantum fluctuations vs classical noise}

\noindent
We now discuss the role of quantum noise in the onset of critical mode competition.
We can see the importance of quantum fluctuations  by comparing the numerical simulations in the beyond-mean-field TW picture with the  same model treated at mean-field (MF) level.
In the latter case, to account for classical variations in the driving strength, we include real-valued spatiotemporal fluctuations to the driving profile.
Details of both models are reported in the Material and Methods section.

 We can calculate the quantity $I_d$ from the outcomes of the two models, which are plotted in Fig.~\ref{fig3}\textbf{(D)} as a blue solid curve (TW) and a pink dashed curve (MF).
Let us now restrict ourselves to the critical region, corresponding to the grey shadowed area in Fig.~\ref{fig3}\textbf{(D)}.
While the TW (blue) curve exhibits a large peak in the vicinity  of the critical point, which results from the mode competition physics discussed in the previous sections, the MF (pink) curve shows clearly an absence of such a peak. 
The peak near the threshold for condensation can be reproduced only by the self-consistent quantum fluctuations accounted within the TWA.
Inspection of the density profiles in the two cases confirms the absence of mode hopping in the latter case.
We demonstrate here that
fluctuations in both density and phase are essential to explain the critical mode hopping;  ``classical'' amplitude fluctuations in the pump strength alone are not sufficient.
While we cannot account for all potential sources of classical fluctuations (e.g. phonons, high energy excitons), we believe that 
they are orders of magnitude weaker than the quantum noise coming from photon decay and pumping (see Supplementary Information
for further discussion).

\

\subsection{Dynamical instabilities at larger pumps}

\noindent

While we have focused so far on the critical region,
it is interesting to also comment on the physics taking place at higher pump powers (i.e., higher polariton densities).
In Fig.~\ref{fig3}\textbf{(D)}, we note the presence of a second peak at $P/P_\mathrm{th}^\mathrm{TW}  \sim 1.92$, i.e. point V.
Our investigation of the polariton and reservoir density profiles suggests that this peak originates from the mechanism of dynamical instability \cite{bobrovska2018dynamical,whittaker2017polariton}.
As discussed in more detail in the Supplementary Information, these can be reproduced by moving to a mean-field formulation of Eq.~\eqref{eq:GPE_pol}-\eqref{eq:GPE_res}, while adding a classical-noise variations to the pump profile (pink dashed curve of Fig.~\ref{fig3}\textbf{(D)}). 
Interestingly, these instabilities are found to be responsible for the large peak in the mean-field (MF) curve too. 
At very large pump powers ($P/P_\mathrm{th}^\mathrm{TW} \sim 3$), the quantity $I_d$ eventually decreases. The density profiles shown in Fig.~\ref{fig2}\textbf{(D)} reveal that at point VI in Fig.~\ref{fig3}\textbf{(D)}, the image difference is due to strong density fluctuations extended over the whole confined region; a different behaviour when compared to the critical mode competition observed at point III.
We expect that such density fluctuations would eventually smooth out ---and resemble more the experimental profiles--- when considering multiple time-windows of the order of tens of microseconds as in the experiment.

\

%
%------------------------------
%-------------------- figure 4
%------------------------------

%%%%%%%%%%%%%%%%%%%%%%%%%%%%%%%%%%%%%%%%%%%%%%%
%%%%%%%%%%%%%%%%%%%%%%%%%%%%%%%%%%%%%%%%%%%%%%%
\section{CONCLUSIONS}

%%%%%%%%%%%%%%%%%%%%%%%%%%%%%%
%%%%%%%%%%%%%%%%%%%%%%%%%%%%%%
%%%%%%%%%%%%%%%%%%%%%%%%%%%%%%
%%%%%%%%%%%%%%%%%%%%%%%%%%%%%%

\noindent
In this work, we investigated experimentally and numerically the density fluctuations in a circularly-confined non-equilibrium Bose-Einstein condensate of polaritons.
As in some experiments with mode hopping in lasers, at the vicinity of the critical region we observe ``telegraphic'' switching with long periods of metastability. 
While the phenomena of mode-hopping is a standard feature in photonic systems such lasers and VCSELs \cite{degen1999transverse,zhang2012towards,deng1997radiation,pereira1998pinning}, in this work for the first time we are able to demonstrate that, in the photonic limit of an interacting polariton fluid, the mode competition arises from the interplay between interactions and quantum noise.
Our results suggest that, analogously to other activation problems, quantum fluctuations are responsible for switching between different states.

In addition, we distinguish between the mode competition at the critical point, and the relaxation mechanisms expected to take place at larger drivings. 
We explore in detail the effects of external perturbations and the role of dynamical instabilities at larger pump powers.
Most importantly, our numerical analysis demonstrates that beyond-mean-field quantum effects are crucial for the onset of mode competition and that it cannot be replicated by a model with just classical noise (i.e., fluctuations in the pump power).
Such findings are determinant for understanding of critical density fluctuations in confined quantum systems. 
Our work paves the way to further investigation of the mechanism of condensate formation in both equilibrium and non-equilibrium settings. 

\

%%%%%%%%%%%%%%%%%%%%%%%%%%%%

%%%%%%%%%%%%%%%%%%%%%%%%%%%%
\section{MATERIAL AND METHODS}

\

\subsection{Sample design}

\noindent
The microcavity used in this work consists of a total of 12 GaAs quantum wells (7 nm thick) with AlAs barriers embedded between two distributed Bragg reflector (DBR) mirrors, which together form an optical cavity (Fig.~\ref{fig1}\textbf{(A)}). The DBRs are made of alternating layers of AlAs and \ce{Al_{0.2}Ga_{0.8}As}. The top DBR is composed of 32 pairs and the bottom DBR is composed of 40 pairs. The quantum wells are in groups of 4, with each group placed at one of the three antinodes of the $3\lambda/2$ cavity. The large number of DBR periods gives the cavity a high Q-factor, resulting in a cavity lifetime of $\sim$135 ps and a polariton life time of $\sim$270 ps at resonance \cite{steger2013long,steger2015slow,steger2012single}. The long cavity life time allows polaritons to propagate over macroscopic distances of up to millimeters. The wavelength of the pump laser was tuned to the second reflective minimum (725 nm), about 113 meV above the lower polariton resonance.

\

%%%%%%%%%%%%%%%%%%%%%%%
%%%%%%%%%%%%%%%%%%%%%%%
\subsection{Numerical model and parameters}

\noindent 
Within the TW approximation~\cite{carusotto2013quantum,estrecho2018single}, the strongly fluctuating polariton field can be described in terms of a stochastic equation of motion coupled to the rate equation for the excitonic reservoir, which read
\begin{align}
 i \hbar d \psi(\textbf{{r}}) &= 
 \begin{aligned}[t] 
  dt\bigg[ 
\left( i \beta - 1 \right) \frac{\hbar^2  \nabla^2}{2 m } 
+ g_\mathrm{c}|{\psi}(\textbf{{r}})|_{{\mathcal{W}}}^2 +g_\mathrm{R} n_\mathrm{R}(\textbf{{r}})+ 
\\
 + \frac{i \hbar }{2} \left( R n_\mathrm{R}(\textbf{{r}}) - \gamma_\mathrm{c} \right) 
\bigg] \psi(\textbf{{r}}) + i \hbar dW_c 
\label{eq:GPE_pol} 
\end{aligned}
\\ 
\frac{d}{dt} n_\mathrm{R}(\textbf{{r}}) &= 
\begin{aligned}[t]
P_r (\textbf{{r}}) - \left( \gamma_\mathrm{R} + R |{\psi}(\textbf{{r}})|_{{\mathcal{W}}}^2 \right) n_\mathrm{R}(\textbf{{r}})
\label{eq:GPE_res}
\end{aligned}
\end{align}
where $\psi(\textbf{r},t)$ is the polariton field, $n_{R}(\textbf{r},t)$ the excitonic reservoir density and $m$ the polariton mass.
$\gamma_\mathrm{c}$ and $\gamma_\mathrm{R}$ define the decay rates of condensed polaritons and the excitonic reservoir respectively.
Following Ref.~\cite{estrecho2018single}, the former term can be estimated as $\gamma_\mathrm{c} = (1-|X|^2) \gamma_\mathrm{ph}$, where $\gamma_\mathrm{ph} = 1/\tau_\mathrm{ph}$ corresponds to the inverse of the photon lifetime and $X$
is the Hopfield coefficient~\cite{Deng2010RMP}.  The excitonic fraction $|X|^2$ is $|X|^2~=~{1}/{2}\left(1+ {\delta}/{\sqrt{4 \hbar^2 \Omega^2 + \delta^2}} \right)$, where $\Omega$ is the Rabi splitting and the detuning $\delta$ corresponds to the difference between the photonic and excitonic energies. 
In Eq.~\eqref{eq:GPE_pol} the Wigner noise accounts for quantum fluctuations and is space- and time- correlated as 
$ \left< dW_c(\textbf{r},t)  dW_c(\textbf{r}^\prime,t) ~\right> =0$, 
$\left< dW_c(\textbf{r},t)  dW_c^*(\textbf{r}^\prime,t)  \right> = (\gamma_\mathrm{c} + R n_\mathrm{R}({\textbf{r}}))/2 \  \delta_{\textbf{r},\textbf{r}^\prime}dt$.
Moreover, the renormalized density $|{\psi}|^2_{\mathcal{W}} \equiv
\left|{\psi} \right|^2 - {1}/{2a^2} $  includes the subtraction of the Wigner commutator contribution, where $a$ is the numerical 2D grid lattice spacing.
The relaxation parameter $\beta$ sets the amount of energy relaxation and increases as the detuning $\delta = E_\mathrm{c}-E_\mathrm{ex}$ increases.
%: $\beta \sim |X|^2$.
The constants $g_\mathrm{c}$ and $g_\mathrm{R}$ are the strengths of polariton-polariton and polariton-reservoir interactions respectively; they can be estimated as $g_\mathrm{c} = g_\mathrm{ex}|X|^4$, $g_\mathrm{R} = g_\mathrm{ex}|X|^2$, with $g_\mathrm{ex}$ the exciton-exciton interaction. The parameter $R = R_0 g_\mathrm{c}/g_\mathrm{R}$ quantifies the stimulated scattering rate of the reservoir excitons into the polariton condensate~\cite{estrecho2018single}.
In order to match the experimental geometry, the pump power profile $P_r(\textbf{r})~=~P ({\gamma_{c}}/{R})  \exp[-\left( {\mathcal{E}}/{2 \sigma^2 } \right)^2]$ is modeled as a tilted elliptical geometry, with a rotated axis of angle $\theta$. $\mathcal{E}$ is the tilted-ellipse equation $\mathcal{E} = Ax^2 +Cy^2 + Bxy -1$ where $A = (\cos^2{\theta})/a^2 + (\sin^2\theta)/b^2$, $B = 2 \sin{\theta} \cos{\theta}$ and $C~=~(\sin^2{\theta})/a^2 + (\cos^2\theta)/b^2$, with $a$ and $b$ the major and minor axis respectively.
The model \eqref{eq:GPE_pol}-\eqref{eq:GPE_res} is solved numerically  adopting an explicit Runge-Kutta method of orders forth and fifth on a two-dimensional numerical grid with $N=128^2$ points with grid-spacing $a = 1.17 \si{\mu m}$. 
We solve the dynamical equations with the following experimental parameters: $m=4.2\times 10^{-5} m_e$ with $m_e$ the electron mass, $\tau_\mathrm{ph} = 135 \si{ps}$, $\gamma_\mathrm{R} = 10^{-3} \mathrm{ps}^{-1}$, $g_\mathrm{c}~=~2~\si{\mu eV \mu m^2}$, $R_0 = 2 \times 10^{-3} \si{\mu m^2 ps^{-1}}$, $\sigma = 1.53 \si{\mu m}$, $b =30 \si{\mu m}$, $a=1.1b$, $\theta = 0.5$.
%%%%%%%%%%%%%%%%%%%%%%%
%%%%%%%%%%%%%%%%%%%%%%%
\subsection{Mean-field with classical fluctuations in the pump}
\noindent
In the main text we discuss the role of quantum fluctuations in the  mode competition  by comparing the results obtained from the TW equations, i.e. {Eq.~(\red{2}) and Eq.~(\red{3})}, with their mean-field approximation and only a classical noise (introduced as fluctuations in the pump power).
% %
The polariton equations of motion in the mean-field approximation read:

\begin{equation}
\begin{aligned}
i \hbar d \psi=d t\left[(i \beta-1) \frac{h^2 \nabla^2}{2 m}+g_c|\psi|^2+g_R n_R+\right. \\
\left.+\frac{i \hbar}{2}\left(R n_R-\gamma_c\right)\right] \psi
\end{aligned}
\label{eq:SGPE_pol}
\end{equation}
\begin{equation}
\begin{aligned}
\frac{d}{d t} n_R=\mathbf{P}_r(\mathbf{r})-\left(\gamma_R+R|\psi|^2\right) n_R
\end{aligned}
\label{eq:SGPE_res}
\end{equation}

To explore the effects of  classical fluctuations in the driving, we introduce spatial-temporal variation in the pump power:
\begin{equation}
\textbf{P}_r(\textbf{{r}}) = \left[ 1 + \chi \ dW(\textbf{r},t) \right]  P_r (\textbf{{r}})
\end{equation}
where dW(\textbf{r},t)  is a zero-mean, real-valued Gaussian noise both in space and time.
This formulation excludes the non-classical correlations 
present in the Truncated Wigner approximation.
Here, the term $\chi$ quantifies the amount of classical fluctuations on top of the driving pump $P_r (\textbf{r})$.

In typical experimental setups, the strength of the classical noise is usually $\approx 1\%$ of the pump intensity.
In our numerical analysis, we push the strength of the noise up to $10\%$ of the total intensity of the pump, corresponding to $\chi = 0.1$. 
Even at these strong classical noise amplitudes, we do not see any peak around the threshold $P_\mathrm{th}^\mathrm{TW} \approx 1$ on the $I_d$ purple curve plotted in Fig.~\ref{fig3}\textbf{(D)}, confirming that the $I_d$ peak measured at criticality is due to the presence of the Wigner quantum noise included in the TW approximation, as discussed in the main text.
We refer the reader to the Supplementary Information for further numerical details.

\section{Acknowledgements} 
We would like to thank A. Opala and M. Matuszewskii for fruitful discussions. 
\section{Funding}
The experimental work at Pittsburgh and sample fabrication at Princeton were supported by the National Science Foundation (Grant No. DMR-2004570). Theoretical analysis was supported by the Engineering and Physical Sciences Research Council (Grant No. EP/V026496/1, EP/S019669/1 and EP/R04399X/1).

\nocite{baboux2018unstable,bobrovska2014stability}
\nocite{snoke2023reanalysis}
\nocite{Bobrovska2015}
\nocite{opala2018}
\nocite{degiorgi2014}
\nocite{Baboux}
\nocite{estrecho2018single}
\nocite{pawelthesis}
\nocite{Baboux}
\nocite{carusotto2013quantum}
%%%%%%%%%%%%%%%%%%%%%%%%% new citations
\nocite{gardiner2004quantum}
\nocite{proukakis2008finite}
\nocite{hartwell2010numerical}
\nocite{snoke2020solid}
\nocite{piermarocchi1996nonequilibrium}
\nocite{myers2018polariton}

\bibliography{biblio}
\clearpage
\title{Critical fluctuations in a confined driven-dissipative quantum condensate}
\maketitle
\beginsupplement

\author{Hassan Alnatah}
\thanks{Address correspondence to: haa108@pitt.edu}
\affiliation{Department of Physics, University of Pittsburgh, 3941 O’Hara Street, Pittsburgh, Pennsylvania 15218, USA}

\author{Paolo Comaron}
\affiliation{Department of Physics and Astronomy, University College London, Gower Street, London, WC1E 6BT, UK}

\author{Shouvik Mukherjee}
\affiliation{Joint Quantum Institute, University of Maryland and National Institute of Standards and Technology, College Park,
Maryland 20742, USA}

\author{Jonathan Beaumariage}
\affiliation{Department of Physics, University of Pittsburgh, 3941 O’Hara Street, Pittsburgh, Pennsylvania 15218, USA}

\author{Loren N. Pfeiffer}
\affiliation{Department of Electrical Engineering, Princeton University, Princeton, New Jersey 08544, USA}

\author{Ken West}
\affiliation{Department of Electrical Engineering, Princeton University, Princeton, New Jersey 08544, USA}

\author{Kirk Baldwin}
\affiliation{Department of Electrical Engineering, Princeton University, Princeton, New Jersey 08544, USA}

\author{Marzena Szymańska}
\affiliation{Department of Physics and Astronomy, University College London, Gower Street, London, WC1E 6BT, UK}

\author{David W. Snoke}
\affiliation{Department of Physics, University of Pittsburgh, 3941 O’Hara Street, Pittsburgh, Pennsylvania 15218, USA}

\section{Switching  Dynamics}
\begin{figure*}
\centering
\includegraphics[width=0.7\textwidth]{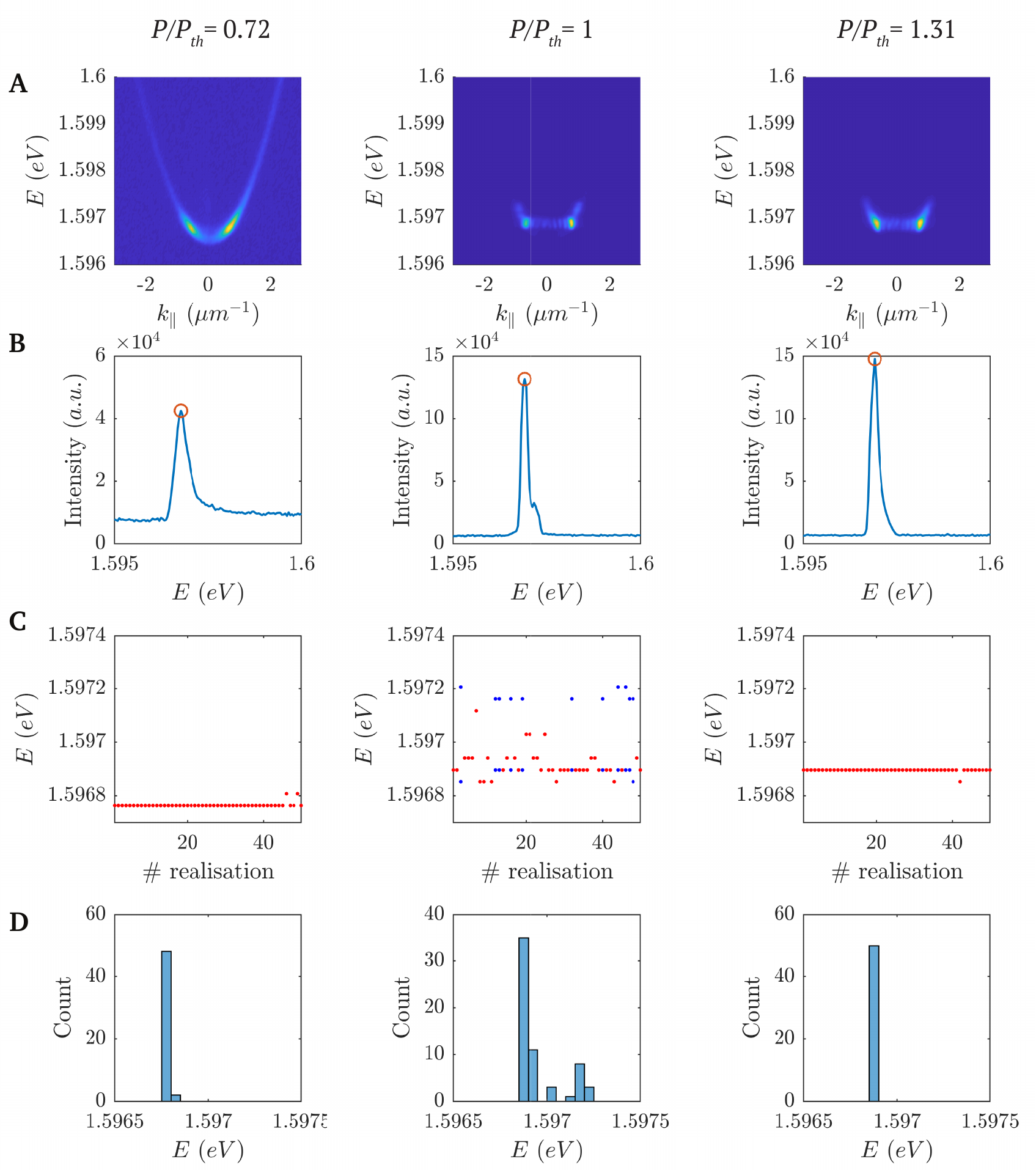}
\caption{\textbf{Experimental energy analysis } 
\textbf{(A)} Representative
$I (k, E)$ snapshots from which the energy maxima were extracted.
\textbf{(B)} $I (E)$ obtained by horizontally summing $I (k, E)$. The red circles illustrate the energy peaks extraction.
\textbf{(C)} Switching dynamics for each snapshot.
\textbf{(D)} A histogram of the occurrence  of modes with given energy. The first column corresponds to below, the second  to near, and the last column to above the condensation threshold. 
}
\label{fig2SM}
\end{figure*}
%
%
%
%------------------------------ fig 3 -------------------
\begin{figure*}
\centering
\includegraphics[width=0.95\textwidth]{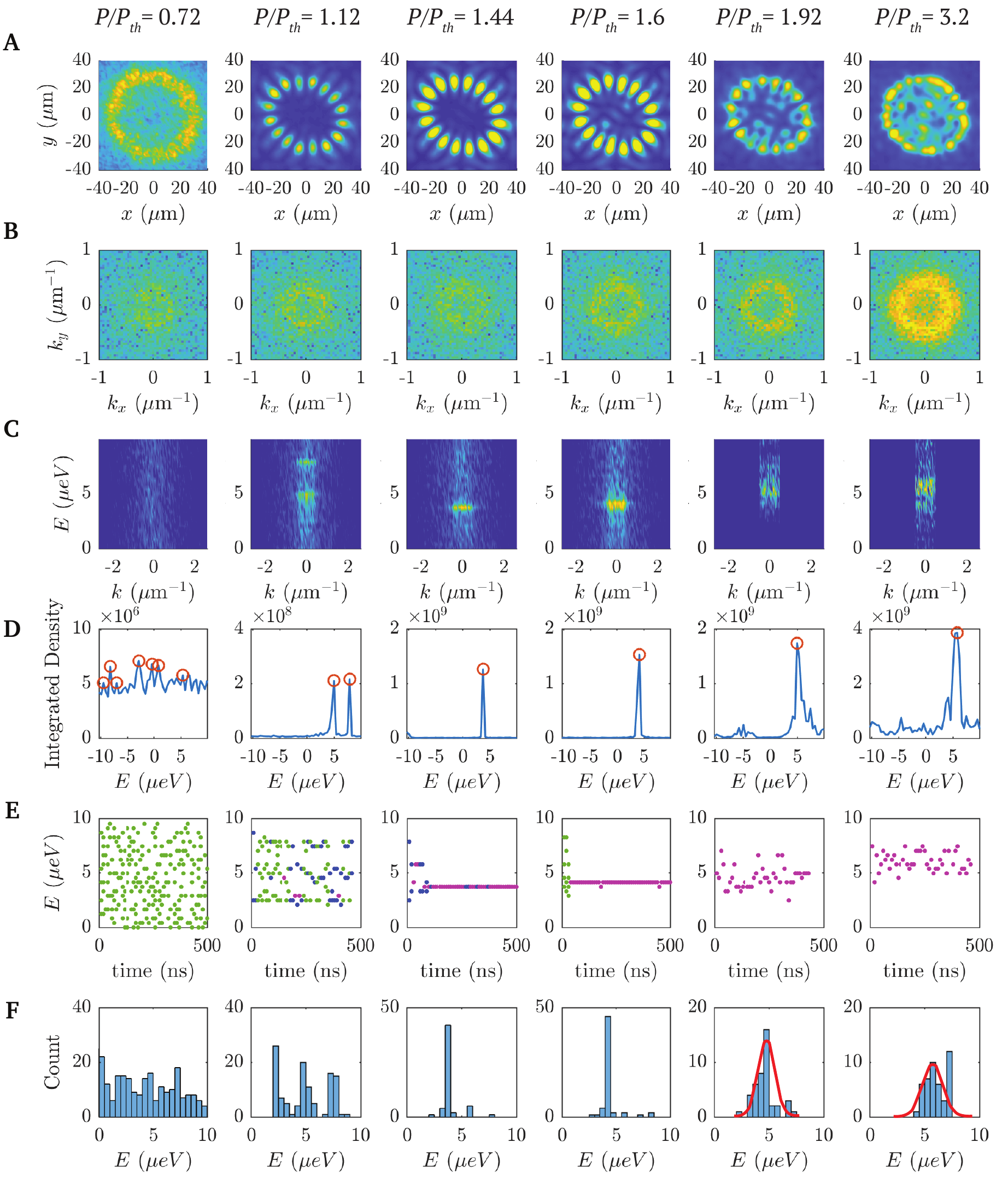}
\caption{\textbf{Theoretical energy analysis.} 
\textbf{(A)} Real space  $\left | \psi(\boldsymbol{r}) \right |^{2}$ and
\textbf{(B)} momentum space $\left | \psi(\boldsymbol{k}) \right |^{2}$ densities. 
\textbf{(C)} $\left |\psi(\textbf{k},E)  \right |^{2}$ from Fourier transforming $\psi = \psi(\textbf{r},t)$ to $\psi = \psi(\textbf{k},\omega)$.
In \textbf{(A-C)} we have integrated over the last 20ns of the dynamics. 
\textbf{(D)} $I(E)$ from integrating  $I(k,E)$  over $k$. The red circles illustrate the energy peaks extraction. 
\textbf{(E)} The energy peaks as a function of time. We periodically sample different temporal windows of $20$ns. Colors of the dots correspond to: single mode (purple); two modes (blue); three and more modes (green).
\textbf{(F)} A histogram of the number of realisations within given energy interval. The red Gaussian curves in the last two histograms are guide to the eye.
}
\label{fig3SM}
\end{figure*}

% \newline
As explained in the main text, experimentally, we did not measure the energy fluctuations in real time. Instead, we took many snapshots of the polariton gas using angle-resolving imaging for each pump power and obtained the mode energy for each snapshot image $I(k,E)$. This was extracted by first summing the image $I(k,E)$ horizontally (i.e integrating over $k$)  to obtain the intensity $I(E)$,  and then numerically estimating the energy for each $I(E)$ peak as shown in Fig.~\ref{fig2SM} \textbf{(B)}. 
The columns in Fig.~\ref{fig2SM} from left to right relate to below, near and above the threshold for condensation (which correspond to points I, II, and III in Fig. 2 in the main text).
We then plot the extracted energy as a function of the snapshot number (Fig.~\ref{fig2SM} \textbf{(C)}). To help visualize the frequency of each mode, we use the points in  Fig.~\ref{fig2SM} \textbf{(C)} to create a histogram (Fig.~\ref{fig2SM} \textbf{(D)}). We find that the polariton gas is multimode near the phase transition and single mode away from the phase transition (i.e above and below the threshold for condensation). 

\par
We follow a similar procedure for the numerical data, which allows us to extract the energy fluctuations (Fig.~\ref{fig3SM}). Fig.~\ref{fig3SM}, columns II, III, V, and VI correspond to points III, IV, V and VI of Fig. 3 in the main text, respectively. These points are discussed in the main text while here we include the full data for completeness. We first integrate the real space solution $\psi(r,t)$ over a $20$ns temporal window. The real space wavefunction $\psi(\boldsymbol{r})$ is then Fourier transformed to $\psi(\boldsymbol{k})$. $\left | \psi(\boldsymbol{r}) \right |^{2}$ and its Fourier transform $\left | \psi(\boldsymbol{k}) \right |^{2}$ are shown in Fig.~\ref{fig3SM}\textbf{(A)} and Fig.~\ref{fig3SM}\textbf{(B)} respectively for the last 20 ns temporal window of the dynamics.
To find the energy resolved spectra, we Fourier transform the wavefunction $\psi = \psi(\textbf{r},t)$ to $\psi = \psi(\textbf{k},\omega)$ over the same $20$ns temporal window. 
We then follow the same method mentioned earlier to extract the energy fluctuations by detecting the energy peaks from the numerical $I(k,E)$ images (see Fig.~\ref{fig3SM}\textbf{(D-F)}). 
We find that below the condensation threshold there is a continuum of energy states (see Fig.~\ref{fig2SM} \textbf{(E)}, column I). Near the critical regime,  the condensate switches between three main states with almost equal distribution (see Fig.~\ref{fig3SM} \textbf{(F)}, column II). Importantly, these critical fluctuations persist over the total simulated time as shown in  Fig.~\ref{fig2SM} \textbf{(E)}, column II.  Well above  threshold,  the condensate undergoes switching  during early dynamics and then settles down to a single mode (Fig.~\ref{fig2SM}\textbf{(E)}, column III). Since the switching happens only during early dynamics, the histogram shows a single dominant mode as illustrated in Fig.~\ref{fig2SM}\textbf{(F)}, column III. 
Fig.~\ref{fig2SM}\textbf{(E)}, columns V and IV show the effect of the dynamical instability  (a term which refers to a spatially fragmented condensate,  caused by effective attractive interactions, mediated by the excitonic reservoir [56,57])  taking place at higher pump powers.
In these cases, the condensate density starts to build up in the centre of the ring, over the whole confined region. What we see then are the spatial fluctuations in the condensate  density  (spatially non-uniform/fragmented condensate) in a single energy mode.   
 This effect is different from the one observed at criticality; a detailed explanation of this phenomenon is given in the next sections.

%%%%%%%%%%%
%%%%%%%%%%%%%%%%%%%%%%%%%%%%%%%%%%%%%%%%%%%
%%%%%%%%%%%%%%%%%%%%%%%%%%%%%%%%%%%%%%%%%%%%%%%%%%%%%%
\section{Statistical Clumping}
\begin{figure*}
\centering
\includegraphics[width=0.5\textwidth]{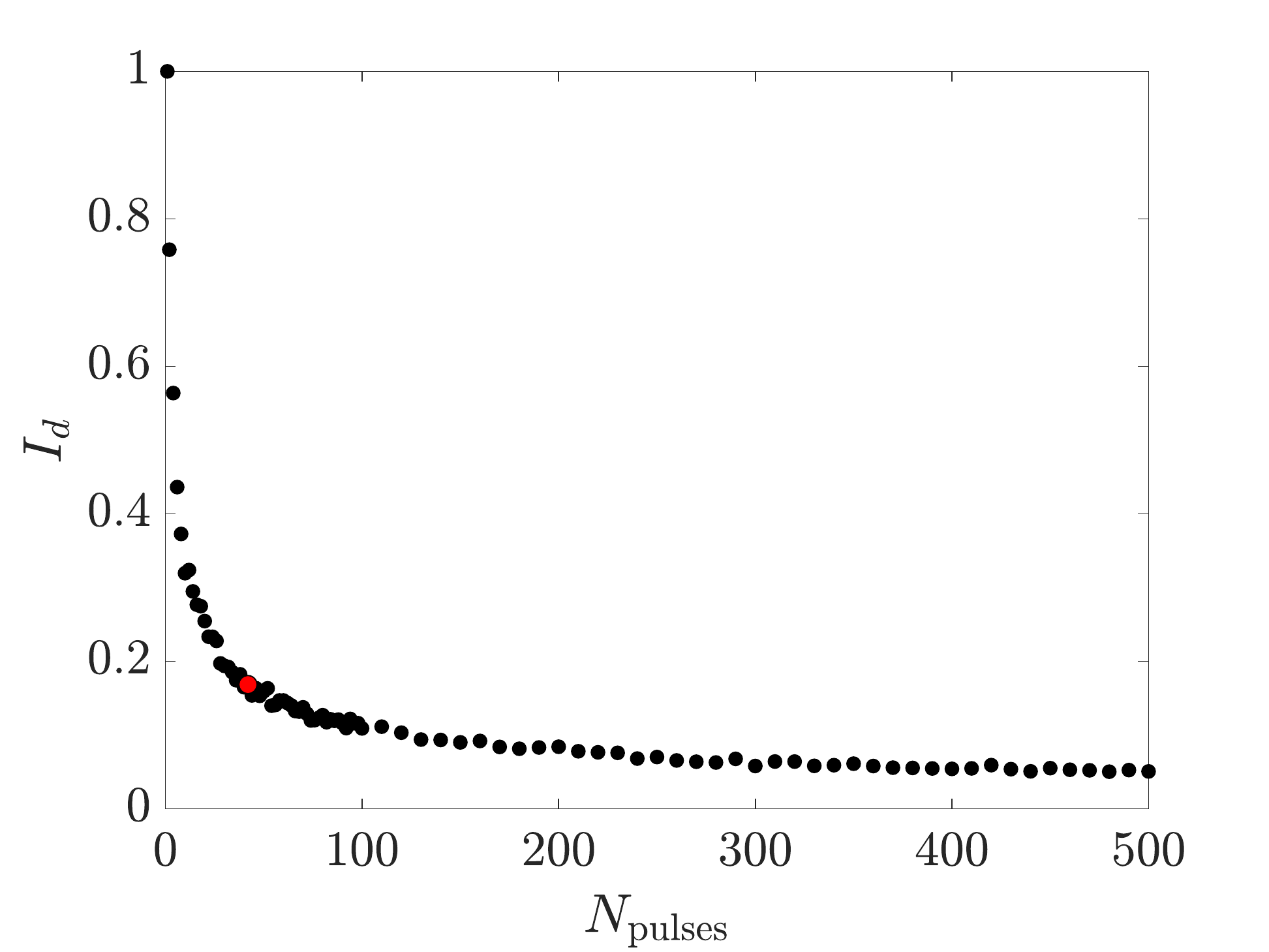}
% \hspace{1.5cm}
% \centering
% \includegraphics[width=0.3\textwidth]{figures/figure1.pdf}
\caption{
\textbf{Statistical clumping of modes.} Normalized image difference by randomly choosing 3 modes $N_{\text{pulses}}$ times, and summing over them. This process is repeated 300 times, each corresponding to a single snapshot. We use Eq.~(\red{1}) in the main text to compute $I_{d}$ . The red circle 
at $N_{\text{pulses}} = 42$ shows the regime we work with experimentally.
% \paolo{increase axis and tick labels up to the text size. the trick is in matlab just reduce the size of the figure. \red{(done)} }
}
\label{clumping}
\end{figure*}

The experimental emission patterns in Fig. \red{2} of the main text are a sum of 42 images from separate pulses, while in the simulations each realization corresponds to a single pulse. In this section, we show that the same mode hopping effect can be reproduced by statistical clumping of certain modes in small samples of 42. To model the statistical clumping, we first start with three modes. We then randomly choose one of these modes $N_{\text{pulse}}$ times and compute the sum of these randomly chosen $N_{\text{pulse}}$ modes. This sum of the $N_{\text{pulse}}$ images is therefore equivalent to a snapshot in the experiment (in the experiment, $N_{\text{pulse}}=42$). Experimentally, we recorded 300
snapshots for each pump power. Therefore, we repeat the above process in the simulations to generate 300 snapshots. We then use Eq.~\red{(1)} from the main text to compute the average image difference of these snapshots. To illustrate the significance of statistical clumping, we vary $N_{\text{pulse}}$ (i.e the number of images that are being summed over) and we plot the image difference as a function of $N$ (see Fig.~\ref{clumping}). As seen in Fig.~\ref{clumping}, statistical clumping of three modes in small sample of 42 (the red circle shown in the plot) shows a noticeable image difference. As expected, when the number $N_{\text{pulse}}$ becomes very large, the image difference tends to zero. However, the experimental regime is in small samples of $N_{\text{pulse}} = 42$ where mode-hopping can still be observed. We find that the image difference $I_{d}$ gets reduced by roughly $0.2$ when $N_{\text{pulse}} = 42$ compared to the case of $N_{\text{pulse}} = 1$. Therefore, the effect of summing over images from separate pulses reduces $I_{d}$ but the overall mode hopping effect can still be observed in a small sample of $N_{\text{pulse}} = 42$.  

{Although the results we have presented in Fig.~2 of the main paper show the condensate in only a single mode, very often we also observe the polariton gas in a linear combination of these modes. However, since we take many snapshots of the condensate, occasionally we also observe it in a single mode even after averaging over 42 chopped pulses. Figure~\ref{multiple_modes} shows an example of polariton condensate in a single model (left and middle panels) and a linear combination of these modes (right panel) for the same pump power $P/P_{th} = 1$. 
\begin{figure*}
\centering
\includegraphics[width=0.9\textwidth]{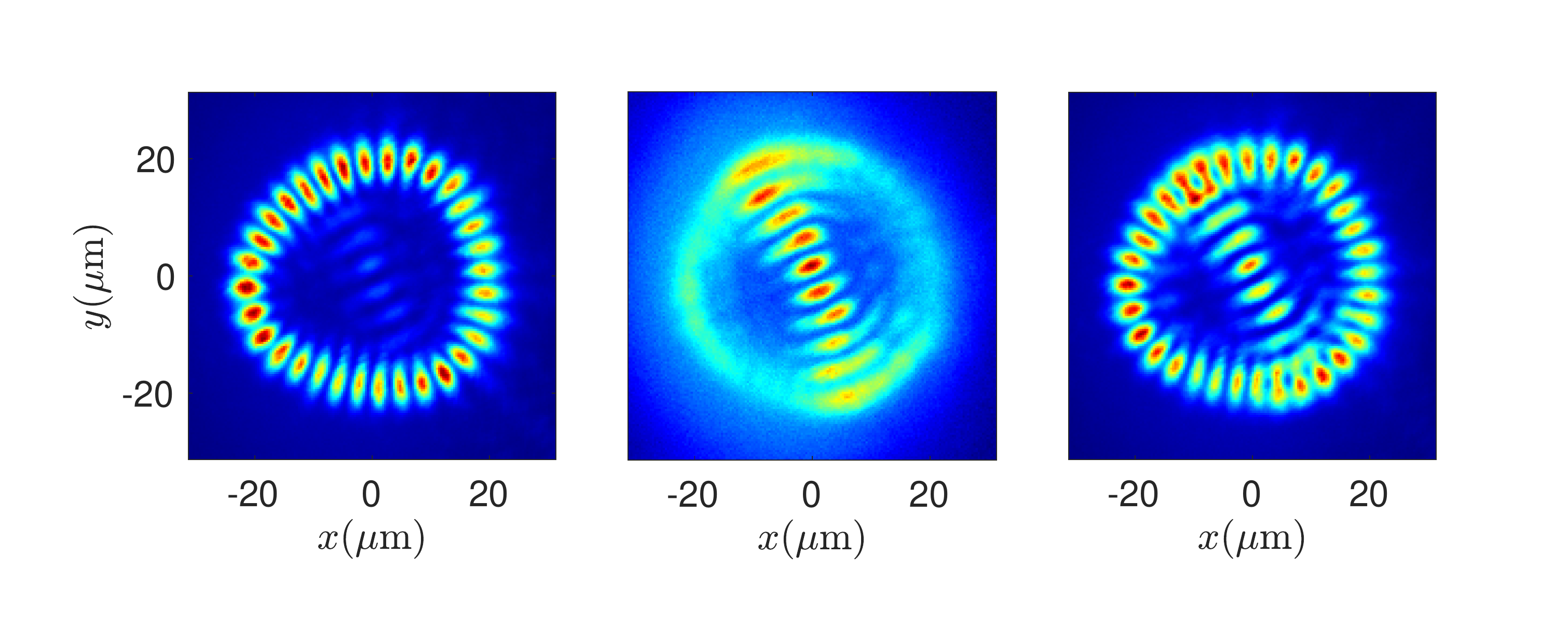}
\caption{
\textbf{Linear combination of modes.} Multiple snpashots of the polarion density profile in real space taken at the threshold power. In the left and middle panels, the condensate is single mode while for the right panel the condensate is in a linear combination of these two modes. 
}
\label{multiple_modes}
\end{figure*}

\section{Energy spacing}
In the experiment, we observe  that energy spacing between the  switching modes of the system is about $200 \;\mu eV$, while in the simulations, the observed energy spacing is $2 \;\mu eV$. In this section, we show that in the simulations the switching happens between states with energy close to the ground state of the system, while experimentally the switching occurs  between higher energy modes. 

\par
To find which of the single particle states the condensates chooses, we solve the Schrodinger equation without nonlinearities in 2D with an annular potential. 
\begin{equation}
-\frac{\hbar^{2}}{2m_{\text{pol}}}\nabla^{2}\psi\left ( \pmb r \right )-V\left ( \pmb r \right )\psi\left ( \pmb r \right ) = E\psi\left ( \pmb r \right )
\label{eq:schrodinger}
\end{equation}
We then compare which of the solutions $\left | \psi_{n,m}\left ( \pmb r \right ) \right |^{2}$ look spatially similar (i.e the same number of pedals or ripples) to the experimental and theoretical real-space images. The eigenstates are calculated using potential of the form $V\left ( \pmb r \right ) = V_{0}\exp\left [-(r-r_{0})^{2}/2\sigma \right ]$, where $r_{0}$ is the radius and $\sigma$ is the width  of the potential ring. We solve Eq.~\eqref{eq:schrodinger} numerically to find $\left | \psi_{n,m}\left ( \pmb r \right ) \right |^{2}$ and $E_{n,m}$. The quantum number $n$ determines the number of nodes in the radial direction, while $m$ determines the nodes around the ring. The number of radial nodes is given by $n-1$ and the number of nodes around the ring is given by $2m$. .

We find that in the experiment the condensate switches between  three main modes $\psi_{(n=1,m=16)}$, $\psi_{(n=5,m=1)}$ and $\psi_{(n=3,m=16)}$, which are shown in Fig. 2, column II of the main text. In the simulations, the switching happens between modes $\psi_{(n=1,m=8)}$, $\psi_{(n=2,m=8)}$ and $\psi_{(n=1,m=1)}$. The energy spacing is roughly proportional to $z\left ( m_{\text{final}}, n_{\text{final}}\right )^{2}-z\left ( m_{\text{initial}}, n_{\text{initial}}\right )^{2}$, suggesting that switching between higher energy modes will give larger energy spacing, where $z\left ( m, n\right )$
is the n-th zero of the regular Bessel function $J_{m}(z)$. Of course, the real-space images of the condensate both in the experiment and the simulations include nonlinearities. However, the solution to the linear Schrodinger equation still allows us to find the quantum number of the  single particle states, which participate in condensation. The states obtained from solving the Schrodinger equation are close to the real space distributions obtained from the experiment and the simulations with only small deviations due to relatively weak polariton interactions.

In our numerical simulations, we find that the modes, which participate in the mode switching  process, depend on various parameters of the model such as the relaxation and interaction strengths. The values of these parameters are not exactly known for semiconductor microcavities [58]. Exploring in principle a large range of possible parameter space is both time-consuming and computationally expensive. We, therefore, focus on reproducing the experiment qualitatively i.e. showing that the mode-hopping is pronounced at the phase boundary and is greatly suppressed away from it.

%%%%%%%%%%%%%%%%%%%%%%%%%%%%%%%%%%%%%%%%%%%%%%%%%%%%%%
%%%%%%%%%%%%%%%%%%%%%%%%%%%%%%%%%%%%%%%%%%%%%%%%%%%%%%
\section{DENSITY CALIBRATION}
In Fig. 2A and Fig. 3A of the paper, we have used the photon counting method to calibrate the density of the polaritons. Photon counting allows us to relate the number of counts on the CCD camera to the number of photons detected. This was done by matching the laser wavelength to the polariton emission wavelength ($776.5\; \mathrm{nm}$). The laser was sent to a mirror at the sample plane, which reflects the laser through the same optical path that was used in the experiment. This reflected beam was then imaged with the CCD camera. The CCD count is proportional to the number of photons, which can be written as:
\begin{equation}
\begin{aligned}
I_{\mathrm{CCD}} = \eta\frac{N_{\mathrm{ph}}}{\Delta t},
\end{aligned}
\end{equation}
where $\Delta t$ is the integration time of the camera, $N_{\mathrm{ph}}$ is the number of photons detected and $\eta$ is the efficiency factor. The number of photons that the camera detects during a time $\Delta t$ is then:
\begin{equation}
\begin{aligned}
N_{\mathrm{ph}} = \frac{P/\Delta t}{hc/ \lambda},
\end{aligned}
\end{equation}
where $P$ is the measured power of the laser, $h$ is Planck constant, $c$ is the speed of light and $\lambda$ is the wavelength of the laser. This then allows us to find a single efficiency factor to convert from CCD counts to number of photons being detected. The efficiency factor is given by:
\begin{equation}
\begin{aligned}
\eta = \frac{hc}{P\lambda} I_{\mathrm{CCD}}.
\end{aligned}
\end{equation}
This efficiency factor was used to calibrate the total density of the polaritons, which is given by:
\begin{equation}
\begin{aligned}
n_{\mathrm{tot}} = \frac{\eta I_{\mathrm{CCD}}\tau}{A_{\mathrm{obs}}},
\end{aligned}
\end{equation}
where $I_{\mathrm{CCD}}$ is the CCD count, $\tau$ is the average radiative lifetime of the polaritons and $A_{\mathrm{obs}}$ is the observed area on the sample from which the light was collected. Since the excitation laser beam was chopped in the experiment with a duty cycle $d = 1.7\%$, then the total number of polaritons is 
\begin{equation}
\begin{aligned}
n_{\mathrm{tot}} = \frac{\eta I_{\mathrm{CCD}}\tau}{A_{\mathrm{obs}}d},
\end{aligned}
\end{equation}
\\
where $\tau \approx \tau_{\mathrm{cav}}/\left|C_{k_{\|}}\right|^2$. Here $\left|C_{k_{\|}}\right|^2$ is the photon fraction and $\tau_{\mathrm{cav}}$ is the cavity lifetime. 

\section{numerical analysis}
%
%--------------------------------------------
%--------------------- fig 4 -----------------------
%--------------------------------------------
\begin{figure*}
\centering
\centering
\raisebox{0.15\height}{\includegraphics[width=\columnwidth]{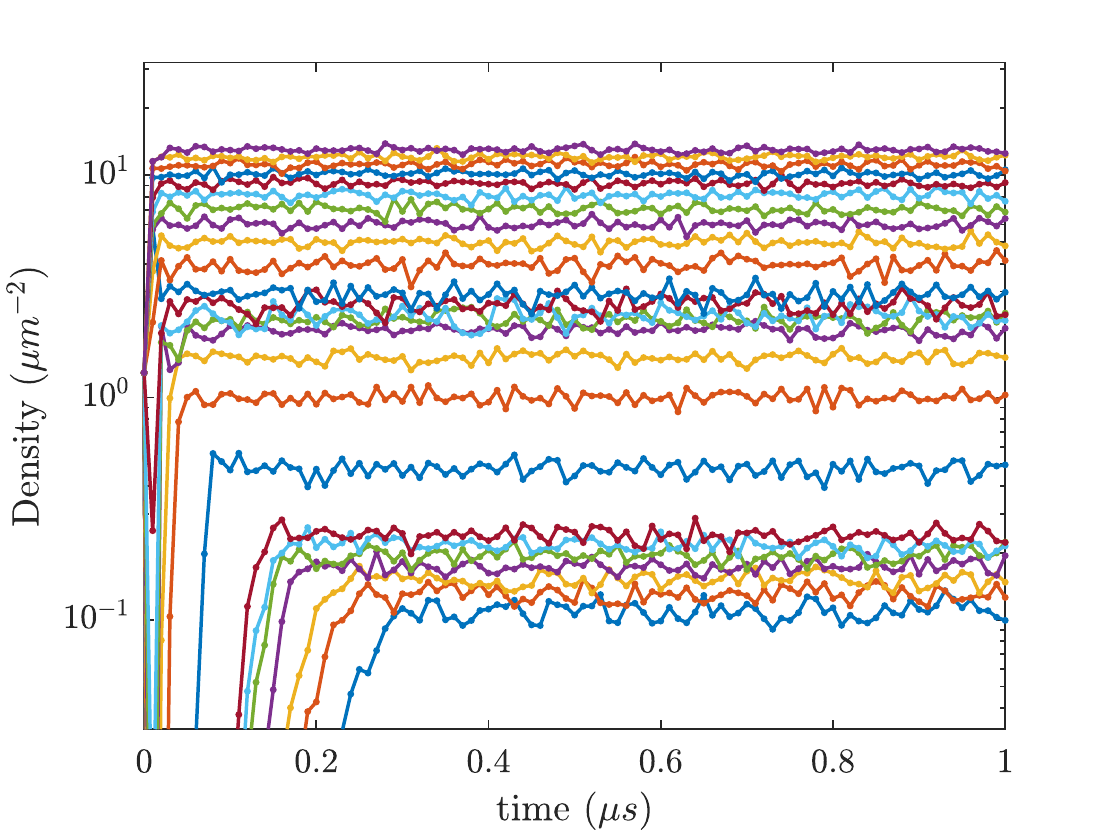}}
\centering
\includegraphics[width=\columnwidth]{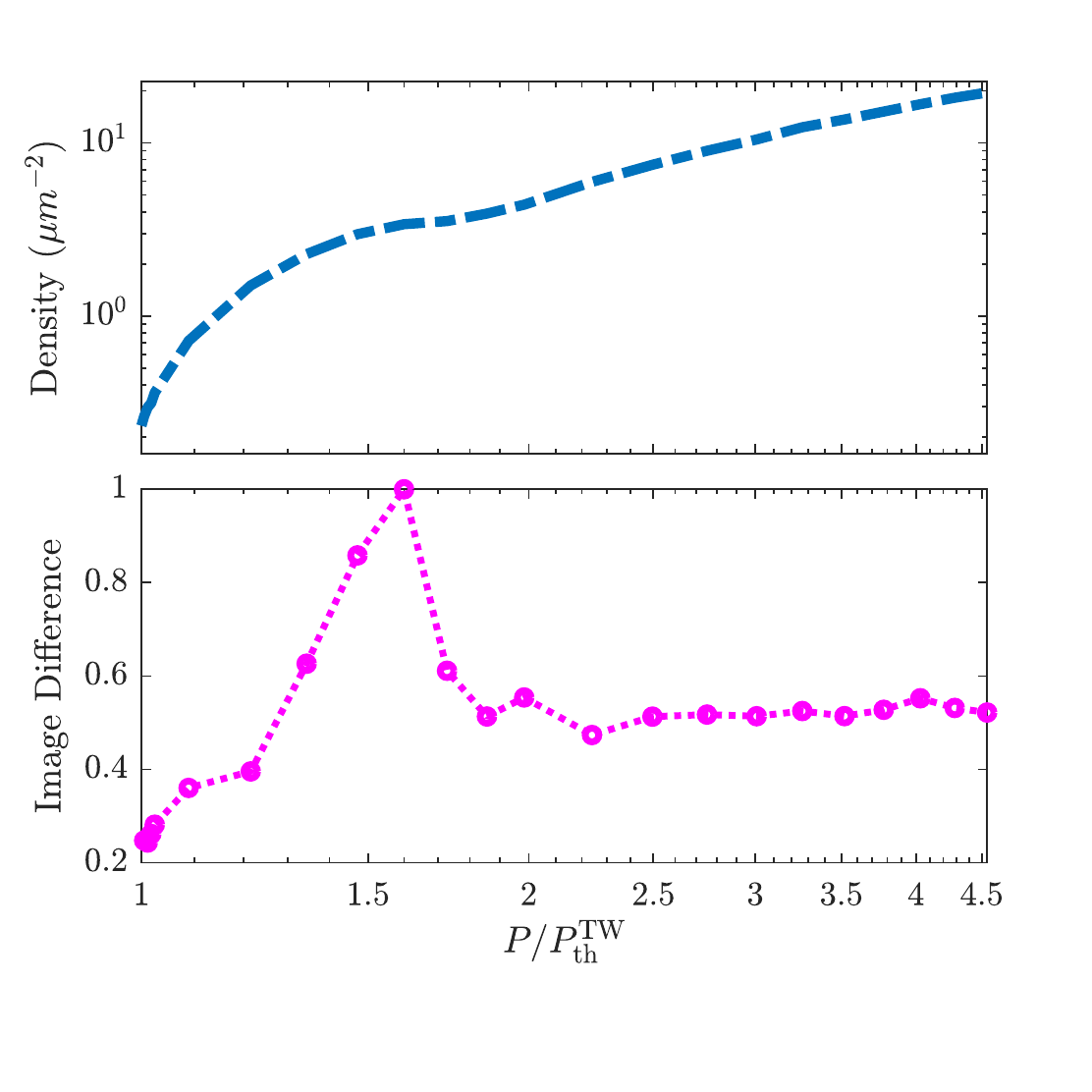}
\caption{\textbf{Mean-field approximation.} \textbf{(left panel)} Time evolution of the total  density $\left< |\psi(t)|^2 \right>_\mathcal{N}$  from Eqs.~(\red{4})-(\red{5}) of the Methods and Material section in the main text
evraged over $\mathcal{N} = 100$ realizations, plotted in a linear-log scale.
Different colors are for different pump powers as in the right panel.
\textbf{(top right panel)} The extracted steady-state densities  as a function of pump power normalised to the condensation threshold $P/P_\mathrm{th}^\mathrm{TW}$ on a log-log scale. The same  as the gray point-dashed curve in Fig.~3 of the main text.
\textbf{(bottom right panel)} Normalized image difference $I_d$  in a log-linear scale obtained by time-integrating the dynamics shwon in the left panel over the interval $0.5\mathrm{\mu s} < T < 1 \mathrm{\mu s}$.
}
\label{fig4SM}
\end{figure*}

In the main text we introduce the numerical model used to simulate the experimental observations, and  discuss the mode switching near the phase transition. In this section, we give more details about the numerical analysis; we first discuss the model and the results from the mean-field equations with noise added to the pump profile (Sec.~\ref{sec:noisy_pump}), we then explain the effects of the system relaxation dynamics on the mode competition (Sec.~\ref{sec:dynamics}), and finally we explore introducing classical external perturbation simulated by a periodic pumping (Sec.~\ref{sec:classical_pert}).

%%%%%%%%%%%%%%%%%%%%%%%%%%%%%%%%%%%%%%%%%%%%%%%%%%%%%%
%%%%%%%%%%%%%%%%%%%%%%%%%%%%%%%%%%%%%%%%%%%%%%%%%%%%%%
\subsection{Numerical details of the mean-field modeling with a noisy pump profile.}
\label{sec:noisy_pump}

In the main paper, we investigate the role of quantum critical fluctuations on the observation of mode-hopping features in the polariton system by comparing different models. 
The  mean-field modeling with a fluctuating pump profile is described in the Methods Section of the main text. In this section, we give further numerical details regarding the simulations of this model.

In Fig.~\ref{fig4SM} we plot the results obtained by solving  Eqs.~\red{(4)-(5)} of the Methods Section in the main text. The nonequilibrium steady state is reached after evolving the system from a random initial condition for $1 \mathrm{\mu s}$. We average over $\mathcal{N} = 100$ realizations. 
The time evolution of the total averaged density $\left< |\psi(t)|^2 \right>_\mathcal{N}$ is plotted in the left panel of Fig.~\ref{fig4SM} for different pump values.
Once the system reaches a nonquilibrium steady state, we extract the average steady-state density shown in the top right panel of Fig.~\ref{fig4SM}.
We then calculate the image difference $I_d$ by means of Eq.~(\red{1}) of the main text and plot it in the bottom right panel of Fig.~\ref{fig4SM}, and in Fig.~3\textbf{(D)} of the main text.

%%%%%%%%%%%%%%%%%%%%%%%%%%%%%%%%%%%%%%%%%%%%%%%%%%%%%%
%%%%%%%%%%%%%%%%%%%%%%%%%%%%%%%%%%%%%%%%%%%%%%%%%%%%%%
\subsection{Effects of dynamical relaxations and instabilities on the mode-hopping.}
\label{sec:dynamics}
We proceed by discussing the effects of the relaxation mechanisms on the mode swithing.
As anticipated in the main text, and shown in Fig.~\ref{fig3SM},
in certain regimes above threshold ($P_\mathrm{th}^\mathrm{TW} \approx 1.5$) the system is seen to undergo swithing between different energy modes during only the early dynamics of the system, before settling down to a single mode. 
Comparison of the time evolution of the polariton and reservoir densities at this particular pump power regime (shown in Fig.~(\red{4}) of the main text) 
suggests that this early-time mode competition originates from the relaxation dynamics in the condensate-reservoir coupled system [48-51,46].
As shown in Fig.~(\red{4}) of the main text, different stochastic realisations (initialized with a different numerical seeds) lead to different dynamics. 
Interestingly, both at criticality and at  early times of the polariton dynamics above threshold, they show “intermittent” oscillations, which resemble a telegraphic behavior. 
Such intermittency may be an effect of introducing the relaxation term $\beta$ in the equations of motions (Eqs.~(\red{2})-(\red{3}) of the main text): preliminary results on the observation of intermittent behavior has been reported in uniform systems [59].
We leave the numerical investigation of this behavior, for the case of confined systems, for future work.

%--------------------------------
%--------------------------------
%-------------- fig 5
%--------------------------------
%--------------------------------
\begin{figure}
\centering
\includegraphics[width=\columnwidth]{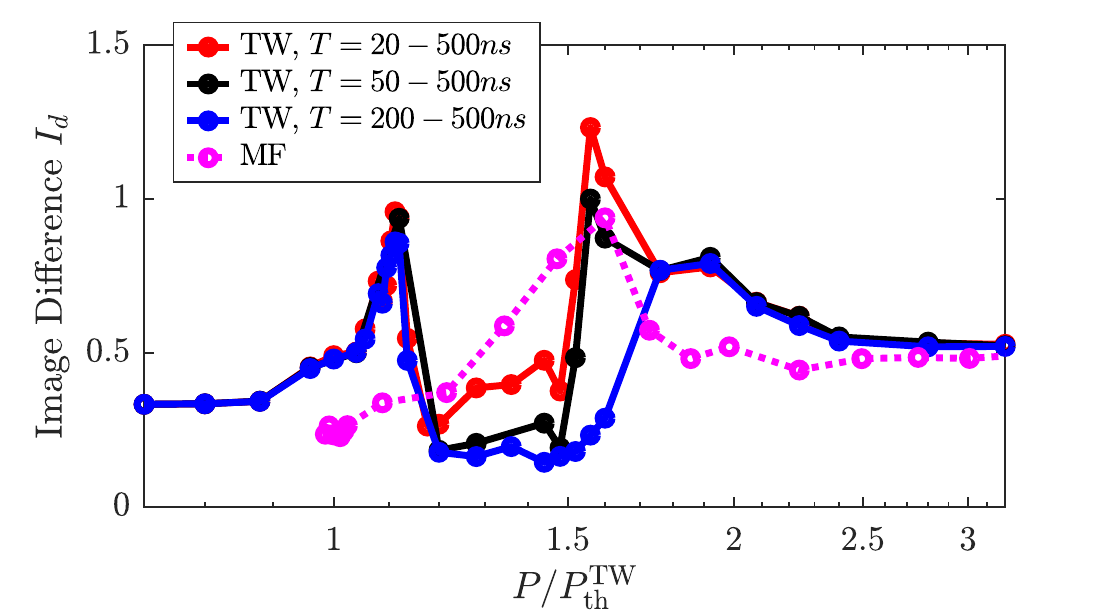}
\caption{\textbf{Image difference.} The image difference $I_d$, calculated within the Truncated-Wigner (Eqs.(\red{2})-(\red{3}) of the main text) marked as solid blue and mean-field approximations (Eqs.(\red{4})-(\red{5}) of the main text) marked as dashed pink line. The same data is shown in Fig.~\red{3}\textbf{(D)} of the main text.
The red and black solid lines correspond to the $I_d$  for the TW, when integrating  over different temporal windows $T$.
}
\label{fig5SM}
\end{figure}
To investigate further the effect of the relaxation dynamics on the mode switching over the whole phase diagram, we extract the quantity $I_d$ for different integration-time windows, and investigate real-space density distributions of the polariton condensate.
In Fig.~\ref{fig5SM}, we plot three different  $I_d$, calculated within the TW approximation, by time-integrating over different temporal windows, from different starting times till the end of the evolution, i.e. $t = 0.5\mu s$.
Specifically, we integrate between: $0.02\mathrm{\mu s} <T<0.5 \mathrm{\mu s}$ (red curve); $0.05\mathrm{\mu s} <T<0.5 \mathrm{\mu s}$ (black curve); $0.2\mathrm{\mu s} <T<0.5 \mathrm{\mu s}$ (blue curve).  
The latter corresponds to the blue curve  Fig.~(3) \textbf{(D)} of the main text.
We note that at $P/P_\mathrm{th}^\mathrm{TW} \approx 1.6$, a large peak appears when considering longer times of integration.
Comparison with the density evolution plots in Fig.~(\red{4}) of the main text suggests that this peak is due to the inclusion of the switching between modes originating from the relaxation mechanisms.

Next, we compare the TW approximation results in Fig.~\ref{fig5SM} with 
the image difference $I_d$, calculated using classical noise added to the pump profile in the MF approximation.
(This is shown in Fig.~\ref{fig5SM} as a pink curve; the same curve is reported in Fig.~\red{3} \textbf{(D)} of the main text.)
Although the MF result does not exhibit a peak at criticality, as discussed in Sec.~\ref{sec:noisy_pump}, it is interesting to note the presence of a maximum  at larger pump powers.
An investigation of the real-space density images along the time evolution  suggests that this is due to the so called modulational instability [51], which is a mean-field effect.   
Attractive interactions between polaritons, caused by the presence of the reservoir, lead to the condensate fragmentation and so fluctuating density, which persist over the whole dynamics as shown in the left panel of Fig.~\ref{fig4SM}.
This indicates a possible relation between the observed MF peak and the peak at $P/P_\mathrm{th}^\mathrm{TW} \gtrsim 1.8$ observed in the TW formulation (blue curve).
% \paolo{\red{add this? -> (To Paolo: I don't think we need to include this but it's up to you.)}
% However, from the analysis above we cannot exclude the presence of other kind  of instabilities; we leave the investigation of these interesting behaviours to future works.}

% \red{DISCUSS THIS //
% This suggests a possible relation between the observed peak and possible relaxation dynamics processes acting at mean field level (pink curve), in the same pump region of the relaxations peaks obtained in the TW formulation (i.e. red and black curve). 
% However, from the analysis above we cannot exclude the presence of other kind  of instabilities, e.g. modulational instabilities~\cite{add}. 
% We leave the investigation of these interesting behaviours to future works. ALSO, LEAVE FIGURE S3-LEFT? questions may arise, REMOVE IT?}

% \paolo{ 
% - we note that the strong mode hopping does not happens exactly at P/Pth = 1, but P/Pth = 1.2. We believe this is due to ..? (discuss this to SM)
% }

\
 
%%%%%%%%%%%%%%%%%%%%%%%%%%%%%%%%%%%%%%%%%%%%%%%%%%%%%%
%%%%%%%%%%%%%%%%%%%%%%%%%%%%%%%%%%%%%%%%%%%%%%%%%%%%%%
\subsection{Modelling classical external perturbation with periodic pumping}
\label{sec:classical_pert}

In the experimental setup, external perturbations e.g. temperature variations or laser modulations may have a large impact on the critical timescales of the system; such perturbations are not included in our model.
In this section we discuss the role of the dynamics of the external pumping on the switching  time-scales of our theoretical modelling.
Specifically, our analysis shows that the time-scale of the mode switching  can be extended up to the whole dynamics (of the order of microseconds) by introducing small periodic modulations of the amplitude of the pump. We model this phenomenon by means of a periodic pump defined as
\begin{equation}
\mathcal{P}_r(\textbf{{r}}) = P (\textbf{{r}}) \left[ 1 + A \cdot \sin(\frac{t}{\tau_{mod}}) \right].
\end{equation}

For our analysis, we choose timescales comparable to the temporal mode-hopping scales of the numerical simulations, namely $\tau_{mod} = 20 \ \mathrm{ns}$.
To probe different regimes of modulations, we vary $A$ in the range $A \in [0, 10^{-2}, 10^{-1} ]$.

Investigation of the real-space density distributions shows that  
a choice of  $A=0.01$ extends the switching time scale to over the whole time-dynamics, namely $0.5 \mu s$. Moreover, comparison of the quantity $I_d$ between modulated and non-modulated cases ($A=0$) shows no appreciable differences ($<10 \%$) in the critical region.
Indeed, the polariton density follows the modulation of the pump, yet the density profiles do not show large changes when compared to the unperturbed case.
To test our results, we do numerics for values up to $A=0.1$.  An investigation of the polariton density distribution shows 
very large oscillations; moreover, the image difference increases substantially compared to the non-oscillating  case.
%%%%%%%%%%%%%%%%%%

\subsection{Effect of static disorder on mode switching}
We have studied the effect of static disorder on mode switching by introducing a time-independent term $V(r)$ to the Hamiltonian of Eq.(2) of the main text. 
We assumed that the random potential has a mean amplitude and root mean square fluctuation given by $\langle V(\mathbf{r})\rangle=0$ and $\sqrt{\left\langle V^2(\mathbf{r})\right\rangle}= 10\; \si{\mu eV}$ respectively. The correlation length of this potential is $2 \; \si{\mu m}$. We have run 50 different stochastic dynamics with the same random static noise profile but with different random noise in the initial wave function. We then averaged over these 50 different stochastic dynamics and repeated the procedure for 50 different random static noise profiles.
\par
We have found that the static disorder enhances the image difference $I_{d}$ by a factor of 4. However, we note that the number of modes that undergo mode hopping is still the same when compared to the case with only quantum fluctuations and no static disorder (i.e the results in the main text shown in Figure 3). Therefore, we conclude that static disorder only introduces density variation of the polariton condensate rather than affecting the number of modes.

%%%%%%%%%%%%%%%%%%%%%%%%%%%%%%%%%%%%%%%%%%%%%%%%%%%%%

\subsection{Convergence Analysis}

In this section, we discuss the numerical convergence in the spatial grid spacing $a$. 
The TWA method, employed in this work, is computationally very efficient and is able to describe quantum fluctuations. However, TWA has a relatively narrow window of applicability: it can be shown [45] that TWA is valid only for lattice spacings greater than the square root of the ratio between interactions and losses: $\gamma_{L P} \gg g_{L P}/a^2$.
This condition essentially sets a lower limit for $a$.
This originates from the derivation of TWA equations, where one assumes the 3rd order derivative in Fokker-Plank equation to be small (at least smaller than the 2nd order term which we keep) so that can be ignored [45]. 

In our convergence and numerical validity tests, we kept the same set of parameters as in the main text and only varied $a$ and pump power $P$. We have run 50 stochastic realizations for different spatial spacing, namely $a = 0.8, 1.02, 1.1, 1.17, 1.55 \;\si{\mu m}$. We have then investigated the time-averaged density profiles for three different pump power values, those corresponding to points III, IV and VI in Fig.~3\textbf{(D)} of the main text.

For the pump power corresponding to point IV (V) in Fig.~3, we find that the condensate is of a single mode (multi-mode) for all $a$ values. 
At the critical point (i.e., the pump power corresponding to point III), we observed that mode switching only happens for $a = 1.1, 1.17 \; \si{\mu m}$, while for the small grid spacing ($a = 0.8\; \si{\mu m}$ and $a = 1.02\; \si{\mu m}$), and for the larger grid spacing ($a = 1.55\; \si{\mu m}$), we observed a single-mode condensate in each stochastic realization. 

For the parameters considered in our work, the TWA condition reads $a \gg \left (g_{L P}/\gamma_{L P}  \right )^{1/2}=0.85\; \si{\mu m}$. 
$a = 0.8$ and $a = 1.02$ can therefore be excluded. 
At large grid spacing (i.e $a \geq 1.55\; \si{\mu m}$), the spatial discretization becomes too coarse for the typical scales of the system. 
Due to a too small high-momentum cut-off ($k_{max} = \pi/a$), momenta contributions that are still significant in the system are suppressed.

\section{Quantum fluctuations vs. classical noise}
Let us first define what we mean by quantum fluctuations. The definition of quantum noise which we use in this study is the same as in the seminal book by Crispin Gardiner and Peter
Zoller [60]. Fluctuations in quantum system arising from interactions
with the external world leading to either drive or dissipation are defined as quantum noise. In contrast, fluctuations
caused by finite temperature are referred to as classical noise. In the truncated Wigner approximation (TWA), quantum fluctuations of the open system manifest themselves as
an additive white noise in space and time present at every time step of the dynamics. This arises from the 2nd order
derivative term of the Fokker-Planck (FP) equations from where TWA is derived. It has to be mentioned that TWA has also been used, especially in the context of cold atoms, to describe finite temperatures [61]. But, as it can be found
in literature, in closed system the 2nd order term is missing in FP equations, and so there is no additive dynamical
noise. Instead, the finite temperature fluctuations are encoded in the initial conditions. The dynamical equations
remain mean-field and the thermal fluctuations are incorporated in the random/noisy initial conditions, where the
observables are computed by averaging over different initial conditions.
\par
Since our system is open (driven/dissipative) and weakly interacting, it is not in thermal equilibrium and the temperature is not defined. If we were to encode some temperature in the initial condition, as in cold atoms, this would have no effect on the steady state as due to continues drive and dissipation (i.e noise at every time step) the influence of any initial conditions will be washed out. Our steady state is initial condition independent.
\par
Based on timescales, we have identified two realistically relevant processes that can induce the mode hopping. The quantum noise acting on photons due to their finite lifetime in the cavity, and the potential variations in the pump power (classical noise). Note that the real-valued noise corresponding to spatio-temporal variations in the pump gives fluctuations in the
density only. There are no fluctuations in the phase of the condensate. What we can say for certain is that we need
fluctuations in the phase to explain the experimental results. Fluctuations in density alone are not sufficient. That is
why we believe that fluctuations in pump power are not the cause of the effects described here. There potentially could
be another source of noise which gives phase fluctuations (external to polariton population) such as phonons or high energy excitons. However, we show in the next sections that interactions with phonons or high energy excitons are orders of magnitude weaker than the  dissipation coming from the finite photon lifetime. 
\subsection{Interaction with phonons}
Modeling thermal fluctuations within the G-P model is a difficult task. To show polariton-phonon interaction being weak, we take a different approach to estimate the polariton-phonon interaction, namely a quantum Boltzmann equation. To simulate the dynamics of the polaritons, we make use of the semiclassical Boltzmann equation, which reads: 
\begin{equation}
\begin{split} 
\frac{\partial n_{\vec{k}}}{\partial t}&=P_{\vec{k}}(t)-\frac{n_{\vec{k}}}{\tau_{\vec{k}}}+\sum_{\vec{k}^{\prime}} W_{\vec{k}^{\prime} \rightarrow \vec{k}}^{(i)}(t)-\sum_{\vec{k}^{\prime}} W_{\vec{k} \rightarrow \vec{k}^{\prime}}^{(i)}(t)
\label{eq:boltz}
\end{split}
\end{equation}
Here  ${n_{\vec{k}}}$ is the occupation number, $\tau_{\vec{k}}$ is the characteristic lifetime, $P_{\vec{k}}$ is the pumping term. The $W^{(i)}$ are the interaction terms for particle-particle collisions. The laser generation is modeled by using a time-independent pump term $P$ that is longer than the total simulated time (i.e c.w pumping). Since the laser generation is nonresonant with energy much greater than the polariton energies, we assumed that the free electrons and holes created in the pump process lead each polariton state being pumped with equal probability. We solve Eq.~\eqref{eq:boltz} numerically by including polariton-phonon interaction and polariton-polariton interaction to find ${n_{\vec{k}}}$. The updated $n_k(t)$ is then used to find the new $W_{\vec{k} \rightarrow \vec{k}^{\prime}}(t)$ until a steady-state distribution is reached. The interactions terms for the polariton-polariton and polariton-phonon interactions are given in Ref [62]. The occupation number of the polartions obtained from Eq.~\eqref{eq:boltz} as a function of their energy is shown in Fig.~\ref{fig1} 
\par
Figure \ref{fig1} shows several important aspects about the exciton-polaritons in general. First, the three regions of the energy distribution have very different properties. At high energy, which corresponds to the excitonic range of the spectrum, there is a thermal tail which fits a Maxwell-Boltzmann distribution since these high energy excitons interact very efficiently with phonons. The temperature of this distribution is very close to the lattice temperature. However, at low energy, which corresponds to the polaritonic region, the polaritons have a nearly flat distribution when only the polariton-phonon interaction is included  and therefore do not have a well defined temperature since they interact very weakly with the lattice. For this reason, we believe that the dissipation coming from the finite photon lifetime is orders of magnitude larger than anything connected with interactions with phonons. When the polariton-polariton interaction is included, the polaritons have a nearly thermal distribution with a temperature well above the lattice temperature.  What we find is that polaritons thermalize with each other and not with the lattice. This is consistent with Quantum Boltzmann solutions previously reported in the literature [62].
\par
\begin{figure}
\centering
\includegraphics[width=0.5\textwidth]{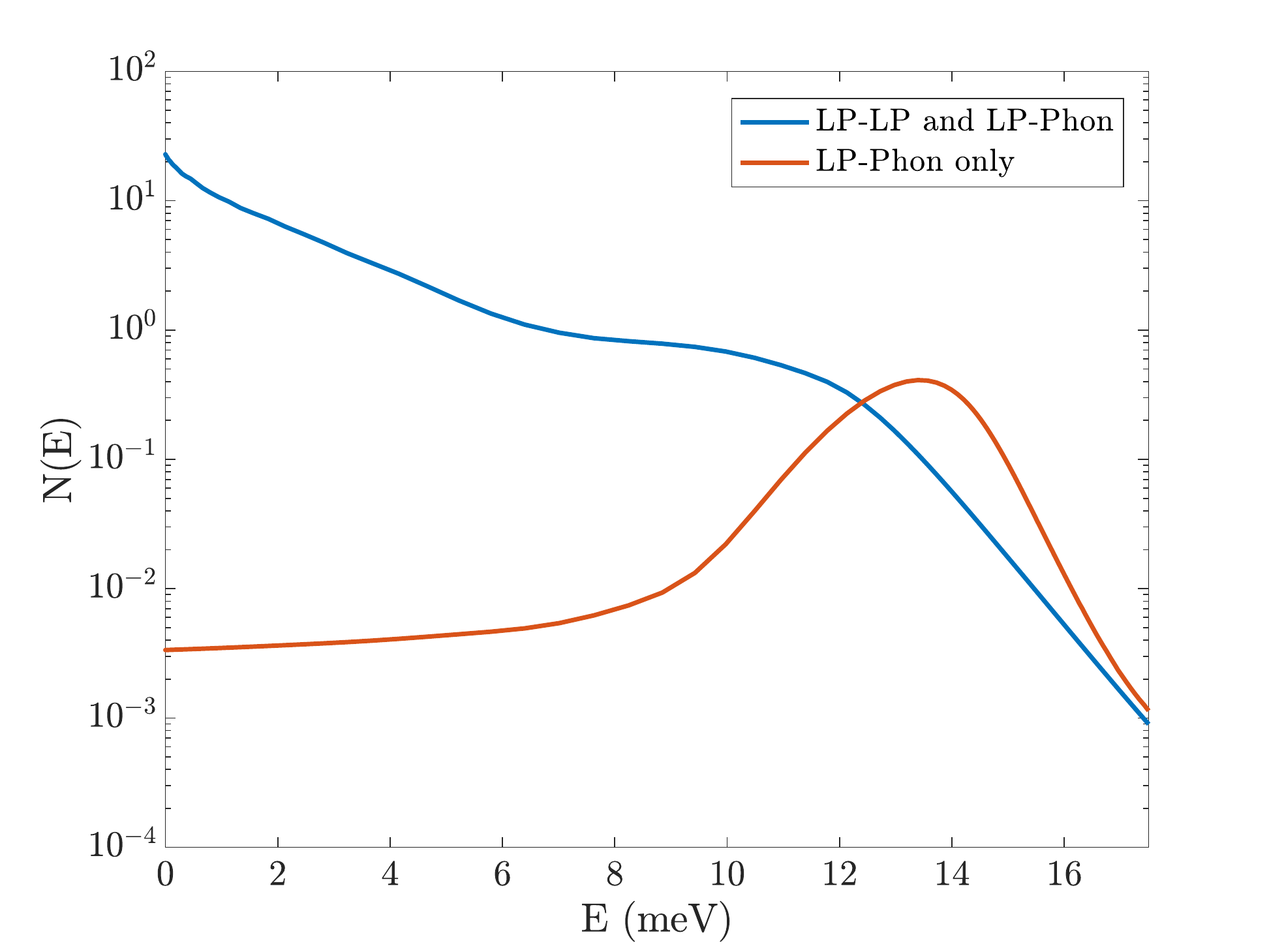}
\caption{The occupation of the lower polaritons as a function of their energy. Red: when only the polariton-phonon interaction is considered. Blue: when both the lower polariton-phonon and polariton-polariton interactions are considered.}.
\label{fig1}
\end{figure}
%%%%%%%%%%%%%
To be more concrete about the strength of this polariton-phonon interaction, we calculate the polariton-phonon out-scattering rate per particle in a momentum state $\mathrm{\vec{k}}$ (Ref. [63], Chapter 4):  
\begin{equation}
\begin{split} 
\begin{aligned}
\begin{gathered}
\begin{aligned}
\frac{1}{\tau_{\vec{k}}} = &\frac{1}{n_{\vec{k}}}\frac{d n_{\vec{k}}}{d t}= \frac{2 \pi}{\hbar} \sum_{\vec{k}_1, \vec{q}_{z}}|M(\vec{k}, \vec{q})|^2\left[n_{\vec{q}}^{\mathrm{phon}}+\frac{1}{2} \pm \frac{1}{2}\right]\\
&\times \delta\left(E_\mathrm{LP}(\vec{k}_1)-E_\mathrm{LP}(\vec{k}) \pm \hbar \nu \vec{q}\right), 
\label{eq:rate}
\end{aligned}
\end{gathered}
\end{aligned}
\end{split}
\end{equation}
where $\pm$ correpsond to phonon emission $(+)$ and phonon absorption $(-)$. Here $n_{\vec{k}}$ is the occupation number of the lower polariton and $n_{\vec{q}}^{\mathrm{phon}}$ is the occupation number of the phonons, which we assume are in thermal equilibrium with $n_{\vec{q}}^{\text {phon }}$ given by the Planck distribution $n_{\vec{q}}^{\text {phon }} = 1/\left (e^{\hbar \omega_{q}/k_{B}T}-1  \right )$.  The polariton-phonon interaction is based on hydrostatic deformation potential, which takes the form [64],
%%%%%%%%%%%%%%%%%%%%%%%%%%%%%%%%%%%%%
\begin{figure}
\centering
\includegraphics[width=0.5\textwidth]{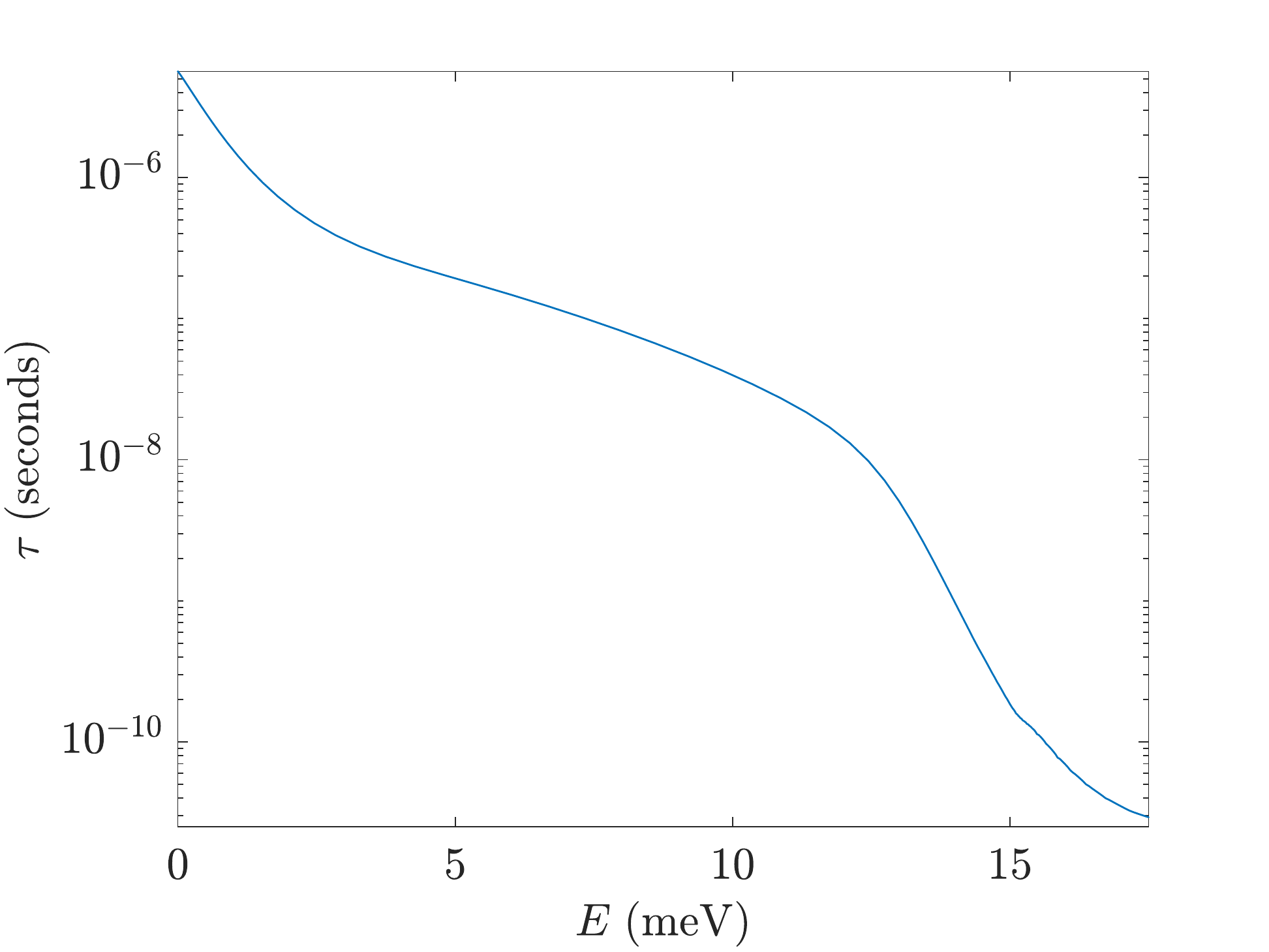}
\caption{The scattering time of polaritons with phonons as a function of the polariton energy calculated from Eq.~\ref{eq:rate}}.
\label{fig2}
\end{figure}
%%%%%%%%%%%%%%%%%%%%%%%%%%%%%%%%%%%%%
\begin{equation}
\begin{split} 
\begin{aligned}
\begin{aligned}
M(\vec{k}, \vec{q})=&  i X_{k} X_{k^{\prime}}\sqrt{\frac{\hbar\left(q_{\|}^2+q_z^2\right)^{1 / 2}}{2 \rho V u}}\\
&\times \left[a_e I_e^{\|}\left((|\vec{q}|) I_e^{\perp}\left(q_z\right)-a_h I_h^{\|}(|\vec{q}|) I_h^{\perp}\left(q_z\right)\right]\right.,
\end{aligned}
\end{aligned}
\end{split}
\end{equation}
where $V$, $\rho$, $u$ are volume, density and longitudinal sound velocity respectively. $X_{\mathrm{k}}$ is the Hopfield coefficient and $a_{e}$ and $a_{h}$ are the deformation coefficients of the conduction and valence band for GaAs respectively. $I_{\mathrm{e}(\mathrm{h})}^{\perp(\|)}$ are the overlap integrals between the exciton and phonon mode. The scattering time is plotted in Fig.~\ref{fig2}. The scattering time in the polaritonic region is in the order of $10^{-6}$ seconds, which is negligible compared to the photon decay which is in the order of a few hundred picoseconds. The scattering time with phonons is four orders of magnitude weaker than the photon decay. In the excitonic region, the scattering time is in the order of tens of picoseconds, which is why they can efficiently thermalize with phonons.
\subsection{High energy excitons}
 In the experimental setup, the high energy excitons stay mostly near vicinity of the pump spot since they are $~10^{4}$ heavier than the polaritons. The diffusion length of these high energy excitons is typically $\leqslant  1\;\mathrm{\mu m}$ while the diameter of the annular trap is $45\;\mathrm{\mu m}$. We therefore expect that the density of these high energy excitons inside the trap where the condensate is forming is nearly zero.  The bottleneck excitons have some polaritonic character and do not exhibit the nearly stationary behavior [65]; this poses an interesting question, however modeling such excitons is difficult and beyond the scope of this work.

%%%%%%%%%%%%%%%%%%%%%%%%%%%%%%%%%%%%%%%%%%%%%%%%%%%%%%
%%%%%%%%%%%%%%%%%%%%%%%%%%%%%%%%%%%%%%%%%%%%%%%%%%%%%%

\end{document}